\documentclass[aps,twocolumn,prx,superscriptaddress,nofootinbib,amsmath,amssymb,floats,floatfix,english]{revtex4-1}
\usepackage{polski}
\usepackage[utf8]{inputenc}
\usepackage[T1]{fontenc}
\usepackage{amsmath}
\usepackage{amssymb}
\usepackage{ifthen}
\usepackage{graphicx}
\usepackage{times}
\usepackage{babel}
\usepackage{color}
\usepackage{bm}
\usepackage[normalem]{ulem}
\usepackage{float}
\usepackage{placeins}
\usepackage{chngcntr}
\usepackage{enumitem}
\usepackage{multirow}

\usepackage{comment}

\newcommand{\ifis}[2]{
\ifthenelse{\equal{#1}{}}{}{#2}
}

\def\be{\begin{equation}}
\def\ee{\end{equation}}
\def\bea{\begin{eqnarray}}
\def\eea{\end{eqnarray}}
\def\bsu{\begin{subequations}}
\def\esu{\end{subequations}}
\def\bi{\begin{itemize}}
\def\ei{\end{itemize}}

\newcommand{\op}[1]{\widehat{#1}}
\newcommand{\dagop}[1]{\widehat{#1}^{\dagger}}
\newcommand{\bo}[1]{{\mathbf{#1}}}
\newcommand{\mc}[1]{{\mathcal{#1}}}
\newcommand{\wt}[1]{{\widetilde{#1}}}
\newcommand{\wb}[1]{{\overline{#1}}}

\newcommand{\nonu}{\nonumber}
\newcommand{\etal}{~\textsl{et al.}}

\newcommand{\bra}[1]{\langle#1\vert}
\newcommand{\ket}[1]{\vert#1\rangle}
\newcommand{\braket}[2]{\langle#1\vert#2\rangle}

\newlength{\templength}

\newenvironment{eq}{\begin{equation}}{\end{equation}}

\newcommand{\eqn}[1]{(\ref{#1})}

\renewcommand{\eq}[2]{\begin{equation}\label{#1}#2\end{equation}}
\newcommand{\eqs}[2]{\begin{subequations}\label{#1}\begin{eqnarray}#2\end{eqnarray}\end{subequations}}
\newcommand{\eqa}[2]{\begin{eqnarray}\label{#1}#2\end{eqnarray}}

\usepackage{lmodern}

\setlist[itemize]{nosep}
\setlist[enumerate]{nosep,nolistsep}

\newcommand{\ve}{\varepsilon}
\newcommand{\cross}{{\sffamily X}}
\newcommand{\tick}{\checkmark}

\newcommand{\tc}[1]{{\large\textcircled{\small{#1}}}}

\begin{document}

\title{A semiclassical field theory that is freed of the ultraviolet catastrophe}

\author{Piotr Deuar}
\affiliation{Institute of Physics, Polish Academy of Sciences, Aleja Lotnik\'ow
32/46, 02-668 Warsaw, Poland}
\email{deuar@ifpan.edu.pl}

\author{Joanna Pietraszewicz}
\affiliation{Institute of Physics, Polish Academy of Sciences, Aleja Lotnik\'ow
32/46, 02-668 Warsaw, Poland}

\date{\today}
\begin{abstract} 
A more accurate semiclassical theory for ultracold gases is derived, in which the occupation of high energy modes is dynamically constrained to the Bose-Einstein distribution. 
This regularized version of the SGPE model preserves the proper nonlinear energy dependence of coupling to the thermal reservoir. 
As a result, inclusion of high energy modes above $k_BT$ does not cause a UV divergence. Instead, the reservoir becomes a constraint on the high energy tails which are included explicitly in the system. Millions of modes can be treated because computational cost scales slowly, like in other semiclassical methods. Implementations in 1d and 3d are presented, 
among them an accurate treatment of the famous case of the quadrupole mode \cite{Jin97}, which had so far eluded satisfactory simulations with any field theory. Our study reveals that observed frequencies and damping of the thermal cloud depended on the experimental signal to noise ratio.
\end{abstract} 

\maketitle 

\section{Introduction}
There is a long list of nonperturbative phenomena in ultracold gases that require semiclassical description. 
The list of applications includes 
quantum turbulence \cite{Berloff02,Wright08}, 
the BKT transition \cite{Bisset09a}, vortex and soliton dynamics \cite{Rooney10,Karpiuk12},
defect formation \cite{Lobo04,Weiler08,Simula14,Fialko15,Liu16},
non-thermal fixed points \cite{Nowak12}, 
the Kibble-Zurek mechanism \cite{Sabbatini12,Swislocki13},
evaporative cooling \cite{Proukakis06,Witkowska11,Liu18}, and more.		
Semiclassical methods (\mbox{c-field}, classical field) \cite{Brewczyk07,Blakie08,Proukakis08,Gardiner03,Sinatra02} are particularly irreplaceable when many
modes are highly occupied or single experimental runs need to be simulated. 
Under such circumstances, only semiclassical approaches remain tractable. 

However, a long standing tough problem there
is that an energy cutoff is needed to deal with the ultraviolet (UV) divergent distribution, 
which emerges in the course  of evolution \cite{Sinatra02,Brewczyk07}.  Effective field theories for 
polaritons \cite{Wouters09,Chiocchetta14}, fermions \cite{Lacroix13,Klimin15}, in cosmology \cite{Opanchuk13,Fialko15,Malkiewicz18}, Yang-Mills theory \cite{Moore97,Tsukiji16}, or nuclear dynamics \cite{Ayik08} share the same issue.
Physically reasonable cutoffs usually occur at energies around $k_BT$. Quantities that have strong contributions from modes around or above this energy become sensitive to the cutoff choice. 
Examples include damping, kinetic energy, and almost any phenomena once temperatures approach the critical temperature. 
While the cutoff can often be optimized to get one observable correct \cite{Witkowska09,Zawitkowski04,Brewczyk07,Cockburn12,Sinatra12,Karpiuk10,Rooney10}, other observables turn out wrong \cite{Pietraszewicz15,Pietraszewicz18a,Pietraszewicz18b}. 
Moreover, a cutoff that gives the correct equation of state and hydrodynamics, makes the resolution at the healing length scale only marginal. That is not enough to properly treat the superfluid defects.
These are the persistent, pervasive, and much hated \emph{cutoff problems} that have often relegated semiclassical simulations to a status of ``only qualitatively accurate''.

The root of the above troubles are differing degrees of freedom between the classical field and the full quantum theory. The former has two degrees of freedom \emph{per mode}, the latter three (or $d$) \emph{per particle}. 
Self-thermalization of an isolated classical field system leads then to an equipartition of $k_BT$ energy per mode, instead of the desired $\tfrac{d}{2}k_BT$ per particle. 
Compared to the Bose-Einstein distribution, an isolated classical field is placing far too much density into the high-energy modes. 
This is what must be changed 
to overcome the cutoff issue at its source.
Some works have managed it in simple systems \cite{Sinatra07,Giorgetti07,Heller09,Heller13,Wouters09}.
However, the unsolved challenge 
is how to do this scalably and independent of favorable symmetries. Then, truly large and general systems can be tackled. 

We provide a route to do so by modifying a variant of the c-field model --- the stochastic Gross-Pitaevskii equation (SGPE) \cite{Stoof99,Gardiner03,Proukakis08}. 
The standard SGPE includes a reservoir that sets the equilibrium temperature of the 
semiclassical field $\phi(\bo{x})$.  
Its implementations to date have imposed a simplified ``classical'' reservoir structure (a Rayleigh-Jeans distribution of reservoir mode occupations) because going beyond this has been difficult for large systems. As a consequence, the high energy components of the field $\phi(\bo{x})$ equilibrated to the usual UV divergent distribution. 
Here, we derive a regularized model and its equations of motion that preserve a \emph{fully quantum description of the constraining reservoir} with Bose-Einstein distributed occupations. 
Moreover, we have found an implementation of this model 
that remains tractable for very large systems ($>10^6$ modes). 
The cutoff can then be moved out to the vacuum to include all the high-energy tails in a convergent, seamless description with the rest of the system. 
This appears to realize a long-held dream in the community to combine the advantageous features of both ZNG and classical field treatments. A large number of interacting low energy modes can be treated non-perturbatively (achieved in classical fields, but not in ZNG which allows only one mode), while the high energy modes couple dynamically to the low energy ones (which appeared in ZNG but not in earlier 
c-field methods). 

As a demonstration, we apply the regularized model to the famous
case of the $m=0$ collective mode \cite{Jin97}, which has resisted all prior attempts at an effective field description for two decades \cite{Bezett09a,Karpiuk10,Straatsma16}. Its correct description has become the standard litmus test for finite temperature field theories of the Bose gas. We will show that the regularized theory is the c-field description that finally passes this test.
Already this first application lets one reach two physical conclusions: that the observed frequencies/damping of thermal clouds depend a lot on the experimental signal-to-noise ratio; and that the low occupied modes can still be usefully represented with a c-field despite a lack of particle discretization.

The paper is structured as follows: 
In Sec.~\ref{RECIPE} we briefly summarize the c-field model and introduce a recipe for improvement.  
The derivation of the regularized SGPE (``rSGPE'') dynamical equations is given in Sec.~\ref{DER}. 
We test its behavior in the single-mode and trapped 1d cases in Sec.~\ref{ANA} to judge its regime of validity. 
Then, our main demonstration -- the description of the $m=0$ collective mode in the JILA experiment \cite{Jin97} is given in Sec.~\ref{JILA}. 
The Appendix 
presents the algorithm developed to run the simulations tractably. More technical aspects of the derivations and data analysis are provided in supplementary material \cite{supp}.

\begin{figure}
\begin{center}
\includegraphics[width=0.49\textwidth]{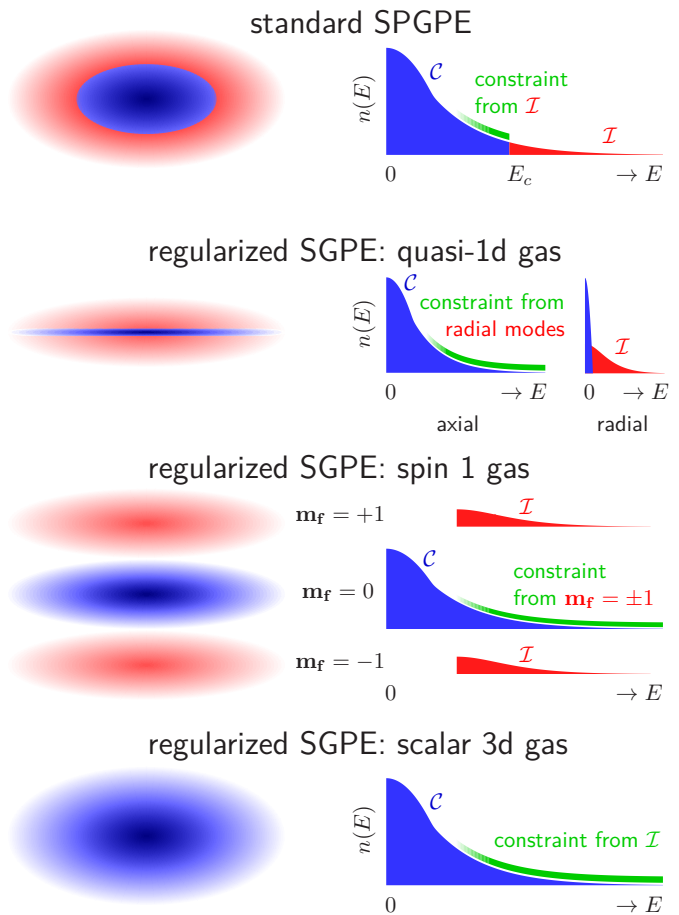}
\end{center}\vspace*{-0.5cm}
\caption{
A schematic illustration of the SPGPE model (top row) and several cases of the regularized model (other rows). 
The right column shows occupations $n$ as a function of mode energy $E$. The left column shows a view of the trapped cloud in x-space.
Colors indicate as follows: 
Red = modes in $\mc{I}$  described by a grand canonical ensemble;  
Blue = modes in $\mc{C}$ described dynamically by a semiclassical c-field $\phi(\bo{x})$ that approximates the quantum field $\op{\phi}(\bo{x})$. 
Green: constraints on the occupations of the $\mc{C}$ modes arising from coupling to the reservoir. 
\label{fig:3x3}}
\end{figure}

\section{Recipe for cutoff elimination}
\label{RECIPE}

\begin{table*}
\begin{tabular}{l||c|c|c|c|c|c|c|c|c|c|}
model		& c-field 				& \multicolumn{2}{c|}{$\mc{C}$--$\mc{I}$ processes}
												& coupling 		& distribution 		& $\mc{C}$ energy		& modes 		& meaning 				& row in\\\cline{3-4}
		& $\phi(\bo{x})$ in $\mc{C}$ 	& growth& scattering	& factor $\mc{G}(E)$ & in $\mc{C}$ tails 	& cutoff $E_c$ 		& in $\mc{I}$ 	& of $\mc{I}, \gamma(\bo{x})$ & Fig.~\ref{fig:3x3} \\
\hline\hline
SPGPE		& \tick				& \tick & $\times$		& linearized	& Rayleigh-Jeans		& $\simeq k_BT+\mu$ 	& $E>E_c$		& reservoir				& 1	\\
rSPGPE	(1d,2d,spin) & \tick			& \tick & $\times$		& full		& Bose-Einstein		& $\to\infty$		& transverse, spin & reservoir 			& 2, 3\\
rSPGPE (scalar 3d)	& \tick			& \tick & $\times$		& full		& Bose-Einstein		& $\to\infty$		& empty		& constraint			& 4\\
\end{tabular}
\caption{
\label{tab:recipe}
A comparison of physical conditions in the SPGPE and regularized (rSPGPE) models. Detail in the text.
}
\end{table*}

\subsection{Existing SPGPE model}
\label{RECIPE1}
Semiclassical methods for the interacting Bose gas are based on a conceptual model that distinguishes two subspaces: $\mc{C}$ and $\mc{I}$.
The split between them is typically made at a cutoff energy $E_c$. 
The low-energy ``coherent''
subspace $\mc{C}$ contains relatively highly occupied and interacting modes which are 
treated non-perturbatively by a complex-valued field $\phi(\bo{x})$.  
These are shown in blue in Fig.~\ref{fig:3x3}. 
In turn, the high energy ``incoherent'' subspace
$\mc{I}$ is comprised of modes with little occupation, and their mutual interactions are neglected. 
Examples, shown in red in Fig.~\ref{fig:3x3}, are the thermal tails and possibly other modes outside of the primary mode space. 

The SPGPE variant of the c-field model reduces the $\mc{I}$ subspace to a static grand canonical ensemble characterized by temperature $T$ and chemical potential $\mu$ \cite{Gardiner03}.
It is assumed that $\mc{I}$ acts as a reservoir for the boson field $\op{\phi}(\bo{x})$ in $\mc{C}$. In order to obtain the standard SPGPE evolution equations\footnote{Shown in Sec.~\ref{S:SPGPE} of the supplementary material \cite{supp}.} a sequence of further assumptions (circled) is made:

\tc{1}: 
Usually only single-particle exchange between $\mc{I}$ and $\mc{C}$ is retained (the so-called ``growth terms''), 
whereas the ``scattering'' terms are left out. 
The dissipation rate of the field 
in $\mc{C}$ turns out to be energy dependent. This dependence is governed by the Gibbs-like factor  
\be\label{GG}
\mc{G}(\hbar\omega) = \exp\left[(\hbar\omega-\mu)/k_BT\right],
\ee
where the frequency $\omega$ is extracted locally from $\op{\phi}(\bo{x})$. 

When energies $\hbar\omega$ are well above $\mu$, mode occupations in $\mc{C}$ are low, and the nonlinear interactions become negligible compared to the dominant reservoir coupling. 
This coupling acts as a constraint on the high energy modes, and their occupations converge to the equilibrium value
\eq{occ}{
N(\hbar \omega) \approx \frac{1}{\mc{G}(\hbar \omega)-1}.
}
This is shown in green in Fig.~\ref{fig:3x3}.

\tc{2}: The energy factor \eqn{GG} has always been linearized 
\eq{linear}{
\mc{G}(\hbar\omega) \to 1 + \frac{\hbar\omega-\mu}{k_BT},
}
to ease the derivation of the stochastic equations and their implementation. This step is a low energy approximation, because it breaks down in the high energy tails regardless of the temperature.

\tc{3}: The underlying operator field $\op{\phi}(\bo{x})$ is replaced by a complex field $\phi(\bo{x})$ in $\mc{C}$.
This is the ``classical field'' approximation and it becomes accurate as mode occupations become large. 
Contrary to a common misconception, \tc{3} is an entirely separate assumption from \tc{2}. This fact will be crucial for improving the theory.

\tc{4}: 
The linearization of \tc{2} requires one to introduce a cutoff in the vicinity of 
\eq{thumb}{
E_c \simeq k_BT + \mu
}
to prevent the UV divergence. At these energies, 
occupations follow the Rayleigh-Jeans law 
\eq{RJ}{
N(\hbar\omega)\to N_{RJ}(\hbar\omega) = \frac{k_BT}{\hbar\omega-\mu}
}
of classical equipartition. Each mode in the tails then adds an energy of $k_BT$, even when occupations decay well below unity. Other semiclassical descriptions such as the projected Gross-Pitaevskii equation (PGPE) \cite{Blakie08,Brewczyk07} or truncated Wigner \cite{Steel98,Sinatra02,FINESS-Book-Ruostekoski} also suffer from the equipartition problem, because of internal ergodic relaxation of the GPE to the same Rayleigh Jeans distribution.

\subsection{Regularized model}
\label{RECIPE2}
We aim to re-derive stochastic equations for the semiclassical field $\phi(\bo{x})$ without making the fateful simplification \tc{2}. 
We continue to assume \tc{1}, and find an alternative route to apply \tc{3}.
The resulting equation constrains the occupations \eqn{occ} to equilibrate to the correct Bose-Einstein distribution. Therefore
 assumption \tc{4} becomes unnecessary, and one can then take the cutoff to any high value $E_c\gg k_BT$ desired. 
In particular, values of $E_c$ at several $k_BT$ reach the asymptotic limit of a cutoff in the vacuum, after which there is no further cutoff dependence. 
The entire system becomes included seamlessly into the $\phi(\bo{x})$ field, like in the bottom row of Fig.~\ref{fig:3x3}.
The high energy tails, which were previously a static reservoir, evolve dynamically. 

The coupling to the reservoir preserves its full energy dependence, and the dissipation rate of the field $\phi(\bo{x})$ can be written as $k_BT\gamma(\bo{x})\left[\mc{G}(\hbar\omega)-1\right]/\hbar$. The prefactor $\gamma(\bo{x})$ 
must take nonzero values to constrain the tails to the right distribution. 
In the regularized model, how $\gamma(\bo{x})$ is chosen depends on whether explicit reservoir modes in $\mc{I}$ are known in the limit $E_c\to\infty$, or not.

Firstly, if there are additional coupled degrees of freedom beyond the primary mode space of $\phi(\bo{x})$, a nominal value of $\gamma(\bo{x})$ can be calculated the same way as for the standard SPGPE. 
For example, in reduced dimensional systems, the transverse modes give a contribution to $\gamma(\bo{x})$ \cite{Bradley15}, which remains unchanged when $E_c\to\infty$. This situation is depicted in the 2nd row of Fig.~\ref{fig:3x3}.
Similarly, in a system with several quasi-spin components and spin exchange \cite{Bradley14}, 
low-occupied modes in higher energy spin states constitute $\mc{I}$, even when the lowest energy spin component is fully contained in $\mc{C}$. 
The spin-1 case is shown in the 3rd row of Fig.~\ref{fig:3x3}. 

In the absence of a clear set of additional modes, the nominal expressions for $\gamma$ used in the past \cite{Bradley08,Rooney12} give a value of zero, once the high energy tails are incorporated into $\mc{C}$. 
Fortunately, there are other physical considerations that 
provide a second route 
and point to which values of $\gamma$ are appropriate.
Namely, the reservoir coupling must be strong enough to hold the modes with energy above $k_BT$ to a Bose-Einstein distribution, instead of the Rayleigh-Jeans one.
Secondly, $\gamma$ must remain small enough to leave the nonlinear low energy modes unconstrained over their natural timescales. A case where $\gamma$ needs to be chosen this way 
is the single-component gas in 3d, shown in the bottom row of Fig.~\ref{fig:3x3}. Appropriate values can be found empirically, or estimated from an ideal gas in the tails.

Note that a better match to experimental dissipation rates has often been obtained using an empirical value of $\gamma(\bo{x})$, several times larger than the nominal one \cite{Bradley08,Proukakis08}. Suspected causes include the ``scattering'' processes omitted by assumption \tc{1} \cite{Rooney12} or various other loss processes neglected in the Hamiltonian. 
The same causes can be physically responsible for nonzero values of $\gamma$ here.
Table~\ref{tab:recipe} summarizes the $\mc{C}-\mc{I}$ coupling in all the variants.

\section{Derivation of the regularized equations}
\label{DER}

\subsection{Stochastic equations}
\label{DER-FULL}


Formally, the Bose field operator $\op{\Psi}(\bo{x})$ for all atoms  
can be expanded over single-particle modes $\op{a}_j$ as
\eq{cf}{
\op{\Psi}(\bo{x}) = \sum_j \op{a}_j \psi_j(\bo{x})
}
with orthogonal basis mode wavefunctions $\psi_j(\bo{x})$ normalized to unity. 
The modes $\op{a}_j$ are most often taken to be plane waves or the harmonic oscillator basis. 
It is convenient to define a projector $\mc{P}_{\mc{C}}$ with matrix elements:
\eq{PCdef}{
P_{\mc{C}}(\bo{x},\bo{x}') = \sum_{j\in\mc{C}} \psi_j(\bo{x})\psi^*_j(\bo{x}'),
}
which extracts the part of the field in $\mc{C}$:
\pagebreak		
\eq{proj}{
\op{\phi}(\bo{x}) = \mc{P}_{\mc{C}}\op{\Psi}(\bo{x}) = \int d^d\bo{x}' P_{\mc{C}}(\bo{x},\bo{x}')\op{\Psi}(\bo{x}') 
= \sum_{j\in\mc{C}} \psi_j(\bo{x})\op{a}_j.
}

Now, a master equation for the reduced density matrix of the $\mc{C}$ subspace, $\op{\rho}_{\mc{C}}={\rm Tr}_{\mc{I}}\left[\op{\rho}\right]$ 
 can be written:

\begin{widetext}
\be\label{mastergeneral}
\frac{\partial\op{\rho}_{\mc{C}}}{\partial t} = -\frac{i}{\hbar}\left[\op{H}_{\mc{C}},\op{\rho}_{\mc{C}}\right] 
+ \int\!d^d\bo{x}\,d^d\bo{x}'\left\{
	\left[\left(G^{(+)}(\bo{u},\bo{v},L_{\mc{C}})\,\mc{G}(\hbar L_{\mc{C}})\circ\op{\phi}(\bo{x})\op{\rho}_{\mc{C}} -  \op{\rho}_{\mc{C}}G^{(+)}(\bo{u},\bo{v},L_{\mc{C}})\circ\op{\phi}(\bo{x})\right)^{\S},\ 
\dagop{\phi}(\bo{x}')\right] 
+ \text{h.c.}\right\}.
\ee
\end{widetext}
This form comes from making assumption \tc{1} to keep only particle exchange with the reservoir. Eq. \eqn{mastergeneral} collects expressions (59), (58), and (37) in Gardiner\etal \cite{Gardiner03}.
The $\bo{x}$ and $\bo{x}'$ are the positions of two particles, while  $\bo{u}=(\bo{x}+\bo{x}')/2$ and $\bo{v}=\bo{x}'-\bo{x}$.
The evolution within the $\mc{C}$ subspace is governed by the Hamiltonian
\begin{equation}\label{HC}
\op{H}_{\mc{C}} = \int\!d^d\bo{x}\ \dagop{\phi}(\bo{x})\left[H_{\rm sp} + \frac{g}{2}\,\dagop{\phi}(\bo{x})\op{\phi}(\bo{x})\right]\op{\phi}(\bo{x})
\end{equation}
with contact inter-particle interactions of strength $g$ and a single-particle energy  $H_{\rm sp}$. Typically 
\eq{Hsp}{
H_{\rm sp} = -\frac{\hbar^2}{2m}\nabla^2 + V(\bo{x}) = \ve + V(\bo{x}),
}
with external potential $V(\bo{x})$ and kinetic energy $\ve$.
  $H_{\rm sp}$ and $\ve$ are linear operations on the field to the right. 

The $G^{(+)}(\bo{u},\bo{v},\omega)$ in the dissipative part of the master equation \eqn{mastergeneral} is a growth rate density of $\op{\phi}$. The corresponding decay rate density $G^{(-)}$ is expressed by $G^{(+)}\mc{G}$. 
Real $G^{(\pm)}$ are assumed and the processes they describe are the transfer of single atoms at energy $\hbar\omega$ between the $\mc{I}$ and $\mc{C}$ subspaces. 
The $G^{(\pm)}$ are understood as operators acting on the quantum field to the right (hence the ``$\circ$'' notation) to extract an appropriate spatial function  $[G^{(\pm)}\circ\op{\phi}\,]$. This function is linear in the mode amplitudes $\op{a_j}$ of the field $\op{\phi}$ and also involves the field's natural constituent frequencies $\omega$. The operator $L_{\mc{C}}=-[\op{H}_{\mc{C}},\op{\phi}(\bo{x})]/\hbar$ 
is used as a shorthand for extracting these frequencies. 
It has the form
\begin{equation}\label{LC}
L_{\mc{C}}\circ\op{\phi}(\bo{x}) = \frac{1}{\hbar}\,\mc{P}_{\mc{C}} \left[ H_{\rm sp}\op{\phi}(\bo{x})  + g\,\dagop{\phi}(\bo{x})\op{\phi}(\bo{x})\op{\phi}(\bo{x}) \right],
\end{equation}
which again acts on the field to the right, and reduces to the GPE frequency in the mean field limit. 
Note that  $L_{\mc{C}}\circ\op{\phi}$ keeps a part nonlinear in $\op{\phi}$.  
Later it will be reduced to terms involving only one factor of $\op{\phi}$ and one of $\dagop{\phi}$, to be consistent with the assumption \tc{1}.  For the time being, we keep account of this matter using the ``\S'' symbol.

The nominal expression for  $G^{(+)}$  in a single-component gas is 
\bea\label{G+}
G^{(+)}(\bo{u},\bo{v},\omega) &=& -\frac{g^2}{(2\pi)^8\hbar^2}\int_{\mc{I}}  d^d\bo{k}_1\,d^d\bo{k}_2\,d^d\bo{k}_3 e^{-i(\bo{k}_1+\bo{k}_2-\bo{k}_3)\cdot\bo{v}}\nonu\\
&\times& F(\bo{u},\bo{k}_1) F(\bo{u},\bo{k}_2)   \left[1+F(\bo{u},\bo{k}_3)\right] \\
&\times& \delta(\omega_{\mc{I}}(\bo{u},\bo{k}_1)+\omega_{\mc{I}}(\bo{u},\bo{k}_2)-\omega_{\mc{I}}(\bo{u},\bo{k}_3)-\omega).\nonu
\eea
The process starts with two particles of momenta $\bo{k}_1$ and $\bo{k}_2$ in $\mc{I}$. It transfers one to $\mc{C}$, while leaving the other particle having momentum $\bo{k}_3$ in $\mc{I}$.
All particles remain in the neighborhood of position $\bo{u}$. 
The
\be\label{homega}
\hbar\omega_{\mc{I}}(\bo{u},\bo{k}) = \frac{\hbar^2|\bo{k}|^2}{2m} + V(\bo{u})
\ee 
are the energies of the particles in $\mc{I}$, while the energy of the transfered particle is $\hbar\omega$, which matches the field $\op{\phi}$. 
$F(\bo{u},\bo{k})$ in \eqn{G+} is the one-particle Wigner function for the $\mc{I}$ particles in the vicinity of $\bo{u}$. It can be written down 
in a simple Bose-Einstein form:
\eq{F}{
F(\bo{u},\bo{k}) = \left[\exp\left(\frac{\hbar\omega_{\mc{I}}(\bo{u},\bo{k})-\mu}{k_BT}\right)-1\right]^{-1}
}
due to the assumption that the particles in $\mc{I}$ are non-interacting and in thermal equilibrium. 


Let us now  proceed with the derivation.
The dissipation and growth rates in the master equation \eqn{mastergeneral} depend in a nonlocal way on $\op{\phi}$ and the quantity $G^{(+)}$, such that the positions of two particles $\bo{x}$ and $\bo{x}'$ need to be taken into account. This convolution of $\op{\phi}$ with $G^{(+)}$ prevents one from obtaining a stochastic equation of only $\op{\phi}$, without storing a huge nonlocal matrix $G^{(+)}$ or explicitly tracking test particles in $\mc{I}$.
However, with a sufficient separation of scales between 
$\op{\phi}$ and the modes in $\mc{I}$, a simplification is doable. 
Suppose the width of $G^{(+)}$ in interparticle distance $\bo{v}$ is narrow compared to the spatial features in $\op{\phi}(\bo{x})$. 
In that case only very closely spaced pairs of atoms will contribute to the product $G^{(+)}\op{\phi}$, and the dissipation of $\op{\phi}(\bo{x})$ will depend only on the local reservoir properties at $\bo{x}$. 
Then, an equivalent local form of the master equation can be obtained.

For the simplification of \eqn{mastergeneral} we will concentrate on an accurate depiction of the dissipation of the low energy modes in $\mc{C}$.
They have been the primary focus of semiclassical treatments. 
The modes in question are those that have high occupation $N(\hbar\omega)\approx k_BT/(\hbar\omega-\mu)\gg1$, which places their energies much closer to $\mu$ than to $\mu+k_BT$. 
Their coupling to all other modes in $\mc{C}$ is well described by the Hamiltonian term, while their fluctuation/dissipation require an accurate representation of $G^{(+)}$ and $\mc{G}$. 
In contrast, for the high energy modes incorporated in $\mc{C}$, it suffices to have just a qualitative rendering of the dissipation rate $G^{(+)}$. The large value of $\mc{G}$ for these modes ensures that they equilibrate rapidly, anyway. 

The required separation of distance scales between the relevant low energy modes in $\mc{C}$ and the modes left in $\mc{I}$ is assured if there is a corresponding separation of energy scales. 
Therefore, it is necessary that all the values of $\hbar\omega_{\mc{I}}$ present in the reservoir are  much larger than the energies of the highly occupied modes in $\mc{C}$, i.e.
\eq{wIcond}{
{\rm min}\left[\hbar\omega_{\mc{I}}\right] \gtrsim \mu+k_BT.
}
Under this condition, $G^{(+)}$ becomes indeed a narrowly peaked function of $\bo{v}=\bo{x}'-\bo{x}$ for all highly occupied modes. As a consequence,
the rendering of $G^{(+)}$ for these modes remains accurate even when one makes the replacement
\eq{spatialG}{
G^{(+)}(\bo{u},\bo{v},\omega)\,\phi(\bo{u}-\bo{v}/2) \approx G^{(+)}(\bo{u},\bo{v},\omega)\,\op{\phi}(\bo{u}).
}
The same conditions on the energy separation also make the replacement 
\eq{G+w0}{
G^{(+)}(\bo{u},\bo{v},\omega)\approx G^{(+)}(\bo{u},\bo{v},0)
}
accurate for the important $\omega$.

In the standard SPGPE, the condition \eqn{wIcond} was usually satisfied implicitly by typical cutoff choices \eqn{thumb}. 
In the regularized model, it is important to explicitly ensure that the chosen subspace $\mc{I}$ satisfies \eqn{wIcond}. 
In particular, the energy gap to other mode spaces should not be smaller than $k_BT$. For example, this concerns the transverse modes or $m_F=\pm1$ spin states in Fig.~\ref{fig:3x3}. 
If the gap is too small, some of the extra modes should be included in $\mc{C}$, so that \eqn{wIcond} is maintained.

Now, applying \eqn{spatialG} and \eqn{G+w0} to \eqn{mastergeneral}, one obtains a reduced local form of the master equation:
\begin{widetext}
\be\label{masterG}
\frac{\partial\op{\rho}_{\mc{C}}}{\partial t} = -\frac{i}{\hbar}\left[\op{H}_{\mc{C}},\op{\rho}_{\mc{C}}\right] 
+  \frac{k_BT}{\hbar}\int\!d^d\bo{x}\, \gamma(\bo{x}) 
	\left\{\left[\left(\mc{G}\circ\op{\phi}(\bo{x})\right)^{\S}\op{\rho}_{\mc{C}} -  \op{\rho}_{\mc{C}}\op{\phi}(\bo{x}),\ 
\dagop{\phi}(\bo{x})\right] 
+ \text{h.c.}\right\}
\ee
\end{widetext}
Here the form $\mc{G}\circ$ operates on the field immediately to its right, such that 
\eq{mcGphi}{
\mc{G}\,\circ\,\op{\phi} = \mc{G}(\hbar L_{\mc{C}})\,\circ\,\op{\phi} = \exp\left[\frac{H_{\rm sp}+g\dagop{\phi}(\bo{x})\op{\phi}(\bo{x})-\mu}{k_BT}\right]\op{\phi}(\bo{x}).
}
The dimensionless coupling strength $\gamma(\bo{x})$, 
discussed in Sec.~\ref{RECIPE2}, is 
\be\label{Gbar}
\gamma(\bo{x}) 
= \frac{\hbar}{k_BT}\ \int d^d\bo{v}\, G^{(+)}\left(\bo{x},\bo{v},0\right). 
\ee

Note, that the equation \eqn{masterG} still 
preserves the full Gibbs factors \eqn{GG} and avoids the somewhat misnamed ``high temperature'' approximation \eqn{linear}.
In fact, temperatures $k_BT\ll\mu$ appear to become treatable, allowing it to cover the majority of phenomena that are of interest 
using a classical field theory.


We now turn to mapping \eqn{masterG} to its equivalent stochastic equations. 
One of the terms is significantly more complex than in the linearized treatment, namely:
\eq{Xdef}{
\op{\mc{X}} = \left[\mc{G}\circ\op{\phi}(\bo{x})\right]^{\S}, 
}
 and requires some care.
It is initially unclear, though, how to best combine the non-commuting operators $\op{\phi}$ and $\dagop{\phi}$ within $\mc{G}$, keeping only one-particle processes, and be rid of the $\S$. 
We will use a positive-P representation 
to deal with this. 
That approach converts the master equation \eqn{masterG} involving  $\op{\phi}$ and $\dagop{\phi}$ to stochastic equations for a pair of corresponding complex fields $\phi(\bo{x})$ and $\wt{\phi}(\bo{x})$.
Effects due to the operator nature of $\op{\phi}$ become encapsulated in the distribution of the complex fields, and $\mc{G}$ can then act on them unambiguously, as detailed in supplementary material, Sec.~\ref{S:PP:NL} \cite{supp}.

In the positive-P representation \cite{Drummond80,Deuar07},
a quantum mechanical density matrix spanning $M$ bosonic modes can be exactly represented as: 
\eq{pprep}{
\op{\rho} = \int d^{2M}\vec{\alpha}\ d^{2M}\vec{\beta}\ P(\vec{\alpha},\vec{\beta})\ \op{\Lambda}(\vec{\alpha},\vec{\beta}).
}
Here $P$ is a probability distribution of simple operator kernels 
\eq{Lambda}{
\op{\Lambda}(\vec{\alpha},\vec{\beta}) = \bigotimes_{j=1}^M \frac{\ket{\alpha_j}\bra{\beta_j^*}}{\braket{\beta_j^*}{\alpha_j}},
}
which are parametrized with continuous variables $\alpha_j$ and $\beta_j$ for the ``ket'' and ``bra'' parts. 
Each $\alpha_j$ ($\beta_j$) is the amplitude of a coherent state $\ket{\alpha_j}$ (\,$\ket{\beta_j}$\,) in the $j$th mode of the system. 
The vectors are $\vec{\alpha}=\{\alpha_1,\dots,\alpha_M\}$, etc. 
The modes $j$ will be all those that constitute the $\mc{C}$ subspace.

The equation of motion for the density matrix can be converted first to a Fokker-Planck equation for $P$, and  next to stochastic equations for fields $\vec{\alpha}$ and $\vec{\beta}$ that are sampled from $P$. 
This proceeds by standard methods \cite{StochMech,QuantumNoise,Drummond80}, and details
of the conversion are given in supplementary material (\ref{S:PP}) \cite{supp}.
In accordance with \eqn{proj}, one can obtain samples of the field $\op{\phi}(\bo{x})$ as
\eq{phipp}{
\phi(\bo{x}) = \sum_{j\in\mc{C}} \alpha_j \psi_j(\bo{x}); \qquad
\wt{\phi}(\bo{x}) = \sum_{j\in\mc{C}} \beta^*_j \psi_j(\bo{x}),
}
for the ``ket'' and ``bra'' states, respectively.
The effect of the procedure is that the equations for the fields $\phi$ and $\wt{\phi}$ turn out relatively simple and efficient, which is a consequence of the local form of the kernel. 
The remaining intractable quantum structure and entanglement between the modes is pushed into the distribution $P$,  which is stochastically sampled. 

Each term in the master equation gives a separate contribution to the equations for the fields $\phi$ and $\wt{\phi}$. 
From the Hamiltonian part one obtains 
\eqs{ppham}{
\frac{d\phi}{dt} &=& -\frac{i}{\hbar}\mc{P}_{\mc{C}}\left[ \left(H_{\rm sp} + g\phi\,\wt{\phi}^* +\sqrt{i\hbar g}\,\xi\,\right)\phi\right]\\
\frac{d\wt{\phi}}{dt} &=& -\frac{i}{\hbar}\mc{P}_{\mc{C}}\left[ \left(H_{\rm sp} + g\wt{\phi}\,\phi^* +\sqrt{i\hbar g}\,\wt{\xi}\,\right)\wt{\phi}\right],
}
while from the diffusive terms {\bf not containing} $\mc{G}$ \cite{Swislocki16}, 
\pagebreak 
\eqs{ppT}{
\frac{d\phi}{dt} &=& \dots + \frac{1}{\hbar}\mc{P}_{\mc{C}}\left[ \gamma k_BT \phi + \sqrt{2\hbar\gamma k_BT}\, \eta \right]\\
\frac{d\wt{\phi}}{dt} &=& \dots + \frac{1}{\hbar}\mc{P}_{\mc{C}}\left[ \gamma k_BT \wt{\phi} + \sqrt{2\hbar\gamma k_BT}\, \eta \right].
} 
The $\bo{x}$ and $t$ dependences
of the fields above have been omitted for brevity. 
The projector $\mc{P}_{\mc{C}}$ on the derivatives keeps the fields within the $\mc{C}$ subspace.
The $\xi(\bo{x},t)$ and $\wt{\xi}(\bo{x},t)$ are two independent real white noise fields with correlations
\eq{xi}{
\langle\xi(\bo{x},t)\xi(\bo{x}',t')\rangle = \delta(\bo{x}-\bo{x}')\delta(t-t'),
}
while $\eta(\bo{x},t)$ is a complex white noise field with $\langle\eta(\bo{x},t)\eta(\bo{x}',t')\rangle =0$ and variance
\eq{eta}{
\langle\eta(\bo{x},t)^*\eta(\bo{x}',t')\rangle = \delta(\bo{x}-\bo{x}')\delta(t-t').
}

The terms in \eqn{masterG} {\bf containing} $\mc{G}$ yield:
\eqs{ppGa}{
\frac{d\phi(\bo{x})}{dt} &=& -\frac{k_BT}{\hbar}\mc{P}_{\mc{C}}\Big\{\gamma(\bo{x})\,\left[\mc{G}\circ\phi(\bo{x})\right]\Big\},\\
\frac{d\wt{\phi}(\bo{x})}{dt} &=& -\frac{k_BT}{\hbar}\mc{P}_{\mc{C}}\Big\{\gamma(\bo{x})\,\left[\mc{G}\circ\wt{\phi}(\bo{x})\right]\Big\}.
}
The form $\mc{G}\circ$ is now acting on complex-valued not operator fields, so there is no longer ambiguity regarding its evaluation. 
Explicitly,
\eqa{mcG}{
\label{newterm}
\mc{G}\circ\phi = \mc{G}(\phi)\,\phi = \exp\left[\frac{H_{\rm sp} + g|\phi(\bo{x})|^2-\mu}{k_BT}\right]\ \phi(\bo{x}).\quad
}
Collecting the terms \eqn{ppham}, \eqn{ppT}, and \eqn{ppGa}, evolution equations fully equivalent to \eqn{masterG} are obtained:

\begin{widetext}
\eqs{rSGPE-pp}{
\hbar\frac{d\phi}{dt}  	&=& \mc{P}_{\mc{C}}\left\{ -i\left(H_{\rm sp} + g\phi\,\wt{\phi}^* +\sqrt{i\hbar g}\,\xi\,\right)\phi 			+ \sqrt{2\gamma k_BT \hbar}\, \eta -\gamma k_BT \left[\mc{G}(\phi)-1\right]\phi 	\right\},\\
\hbar\frac{d\wt{\phi}}{dt} 	&=& \mc{P}_{\mc{C}}\left\{ -i\left(H_{\rm sp} + g\phi\,\wt{\phi}^* +\sqrt{i\hbar g}\,\wt{\xi}\,\right)\wt{\phi} 	+ \sqrt{2\gamma k_BT \hbar}\, \eta -\gamma k_BT \left[\mc{G}(\wt{\phi})-1\right]\wt{\phi} \right\}.
}
\end{widetext}

\subsection{Reduction to a classical field}
\label{PP-CF}

To have tractable long-time evolution, one has to make a reduction to a semiclassical field. 
Terms containing the real noises $\xi$ and $\wt{\xi}$ eventually would be
responsible for instability and dynamical quantum fluctuation effects beyond the classical field picture. They should be removed \cite{Gilchrist97,Deuar06b}. 
Notably, reasonable initial states (thermal, coherent, vacuum, \dots) have $P$ distributions with the nice feature that $\phi(\bo{x})=\wt{\phi}(\bo{x})$ for all samples. Therefore, in the absence of the $\xi$ and $\wt{\xi}$ noises, the values of $\phi(t)$ and $\wt{\phi}(t)$ become identical. 
Hence, setting
\eq{setc}{
\xi=\wt{\xi}=0\qquad;\qquad\wt{\phi}=\phi
}
directly and cleanly imposes the classical field approximation. We have then, the regularized SPGPE (rSPGPE):
\eqa{rSPGPE}{
\hbar\frac{d\phi}{dt}  	&=& \mc{P}_{\mc{C}}\bigg\{ -i\left[\,H_{\rm sp} + g|\phi|^2\, \right]\phi 	+ \sqrt{2\gamma k_BT \hbar}\, \eta \\
&&-\gamma k_BT \left[\,\exp\left(\frac{H_{\rm sp} + g|\phi|^2-\mu}{k_BT}\right)-1\,\right]\phi 	\bigg\}.\nonu
}
The details of the projection $\mc{P}_{\mc{C}}$ become irrelevant once the energy cutoff $E_c$ is taken beyond all appreciably occupied modes. 
Accordingly, we will implement the numerically simplest case: a plane wave basis with momentum cutoff $k_{\rm max}=\pi/\Delta x$, $\ve_{\rm cut} = \hbar^2 k_{\rm max}^2/2m$ set by the numerical grid. 
Explicit projection is discarded, giving the regularized SGPE (rSGPE):
\begin{widetext}
\eq{rSGPE}{
\hbar\frac{d\phi}{dt}  	=  -i\big[\,H_{\rm sp} + g|\phi|^2\, \big]\phi 	+ \sqrt{2\gamma k_BT \hbar}\, \eta -\gamma k_BT \left[\,\exp\left(\frac{H_{\rm sp} + g|\phi|^2-\mu}{k_BT}\right)-1\,\right]\phi .
}
\end{widetext}
This equation 
is significantly simpler to implement than \eqn{rSPGPE} and has a more pronounceable acronym. 
Although simpler does not mean 
trivial, since the dissipation term  contains an exponential 
of non-commuting quantities:  
the kinetic energy $\ve$ vs the potential $V(\bo{x})$ and the interaction term $g|\phi(\bo{x},t)|^2$.

For a small system, the exponential could be dealt with exactly using a matrix representation (see e.g. \cite{Wouters09}) or a diagonalization of $H_{\rm sp}$. 
In large systems, such as most in 2d or 3d, this is impractical. 
In an interacting system, diagonalization and matrices would have to be re-calculated at each time step, making the whole procedure even more prohibitive computationally. 
Furthermore, because the dissipation is large, it seems that small timesteps are needed.
These computational matters have presumably contributed to why related stochastic equations 
have not been utilized in the past. 
For example,  Duine and Stoof \cite{Duine01} derived a Fokker-Planck equation (eq. (32) therein)  with similar exponential Gibbs factors, but noted that ``its solution is very difficult numerically'' and proceeded to linearize in temperature and derive the standard SGPE.

The Appendix gives details of the 
algorithm we developed to make the integration of 
\eqn{rSGPE} tractable.
The resulting computational effort remains comparable to that of the plain SGPE. The required time step $\Delta t$ stays of a similar size despite the much larger dissipation, and the  
number of operations per time step scales with the number of modes like $M\log M$.

\section{Analysis and testing}
\label{ANA}

\begin{figure}
\begin{center}
\includegraphics[width=\columnwidth]{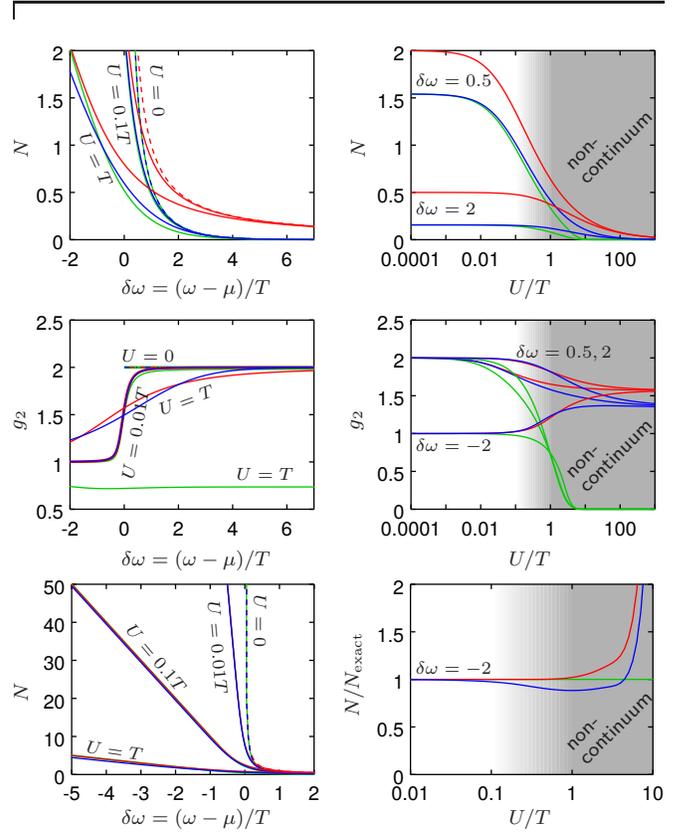}
\end{center}
\vspace*{-0.5cm}
\caption{
Characteristics of the single-mode equilibrium state. 
Left column: dependence on single-particle energy.
Right column: dependence on the ``energy band'' spacing $U/T$.   
Top row: occupation $N=\langle\dagop{a}\op{a}\rangle$;
Middle row: $g_2$;
Bottom row: behavior in the ``Thomas-Fermi'' regime of $\omega<\mu$.
The less relevant non-continuum Mott-insulator-like regime is in gray.
Green: exact quantum mechanics, red: standard classical field, blue: regularized SGPE. $U=0$ case shown dashed.
}
\label{fig:1mode}
\end{figure}

\subsection{Single mode analysis}
\label{1MODE}

Basic information about a tested method 
is given by 
analysis of a single mode system.
When a single mode is rendered incorrectly, the same failure will occur in a many mode system for similar regimes. 
Conversely, where a local one-mode model works well, one expects that at least \emph{local} observables will be described accurately. 

Let the mode in question be a small volume $\Delta v$ around a particular point $\bo{x}$ in the gas, and  
$ \op{a} \approx \op{\phi}(\bo{x})\sqrt{\Delta v}, 
\alpha \approx \phi(\bo{x})\sqrt{\Delta v}$.
 The Hamiltonian is  
\eq{H1}{
\op{H}_1 =\hbar\omega\,\dagop{a}\op{a} + \frac{U}{2}\dagop{a}{}^2\op{a}^2
}
with a local single particle energy $\hbar\omega$ 
and Bose-Hubbard-like interaction energy $U~=~ \frac{g}{\Delta v}$.
Taking $\hbar=k_B=1$, 
the equation of motion for the mode amplitude $\alpha$ is then:
\eq{motion1}{
\frac{d\alpha}{dt}  	= -i\left[\omega + U|\alpha|^2\, \right]\alpha	 -\Gamma_1\, \alpha\ + \sqrt{2\gamma T}\, \eta_1(t)
}
where $\eta_1(t)$ is a complex noise of variance $1/\Delta t$, and 
\eq{motion2}{
\Gamma_1=
\left\{\begin{array}{cl}
\gamma T \left(\,e^{\frac{\omega-\mu + U|\alpha|^2}{T}}-1\,\right) 	&\text{ for rSGPE,} \\
\gamma\left(\,\omega-\mu~+~U|\alpha|^2\,\right) 				&\text{ for SGPE.}
\end{array}\right.
}

The thermal distribution of the standard classical field is given in terms of $n=|\alpha|^2$ by 
\eq{Pcfield}{
P_{\rm cf}(\alpha) = {\rm const}\times \exp\left[-\frac{1}{T}\left(\omega-\mu+\frac{Un}{2}\right)n\right].
}
For the rSGPE the stationary distribution can also be obtained: 
\eq{Preg}{
P(\alpha) = {\rm const}\times \exp\left[-\frac{Te^{(\omega-\mu)/T}}{U}\left(e^{\frac{nU}{T}}-1\right) + n\right],
}
remarkably, 
even in the interacting case (supplementary material~\ref{S:PP:1MODE} \cite{supp}).

When $U=0$, this distribution becomes a Gaussian 
$P_{U=0}(\alpha) \propto \exp\left[-\left(e^{\frac{\omega-\mu}{T}}-1\right)n\right]$, 
whose average mode occupation exactly agrees with the quantum value, i.e.
\eq{NU=0}{
\langle n\rangle_{U=0} = \frac{\int n P_{U=0}(\alpha)\ d^2\alpha}{ \int P_{U=0}(\alpha)\ d^2\alpha} = \frac{1}{e^\frac{\omega-\mu}{T}-1} = N_{BE}(\omega).
}
In contrast, the standard classical field distribution \eqn{Pcfield} gives $\langle n\rangle_{U=0}=T/(\omega-\mu)$. 

Still, the rSGPE distribution is not a full quantum description of the mode, because  
higher order moments may not agree. However, the on-site two-particle correlation
\eq{g2}{
g_2 = \frac{\langle\dagop{a}{}^2\op{a}^2\rangle}{\langle\dagop{a}\op{a}\rangle}, 
}
is correctly reproduced by its classical field estimate $g_2^{\rm est}=\langle n^2\rangle / \langle n\rangle^2=2$ in the $U=0$ limit. This is
because the distribution $P(\alpha)$ is Gaussian. 

The observable predictions obtained with the rSGPE are shown in Fig.~\ref{fig:1mode} in blue, as compared to standard c-\\fields (red) and exact quantum values (green). The gray background in the figure corresponds to the regime  $U\gtrsim T$ that appears when the mode volume $\Delta v$ is too small. It is not applicable to continuum systems and leads to bogus physical consequences: 
spurious energy bands in the spectrum and a Mott-insulator-like state. 
Several clear points emerge from Fig~\ref{fig:1mode}:
\begin{itemize}
\item The incorrect occupations of the standard approach are greatly improved by the regularization for all $U$. They are essentially exact for $U\lesssim T$ (top row). 
\item The fluctuations $g_2$ (center row) switch between being accurate for both SGPE and rSGPE in the usual continuum regime of $U\lesssim 0.1T$, to being strongly incorrect once the gray non-continuum regime is reached.  
In particular, 
the $U=0.01T$ line, which lies among typical physical parameters for a dilute gas is well reproduced, even though it has a nontrivial dependence on $\omega$.
\item In the ``Thomas-Fermi'' regime $\omega<\mu$ in which an interaction-induced condensate or quasicondensate forms (bottom row), both standard and regularized c-fields give the same correct results.
\end{itemize}

\subsection{Test cases in 1d}
\label{1D}

\begin{figure}
\begin{center}
\includegraphics[width=0.8\columnwidth]{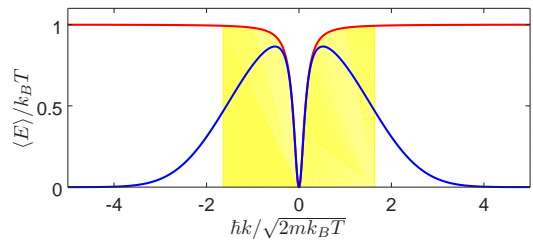}
\end{center}
\vspace*{-0.5cm}
\caption{
Energy per mode $\langle E\rangle=\hbar\omega\langle n\rangle_{U=0}$ in the uniform ideal gas (grand canonical ensemble with $\mu=-0.01k_BT$). Red: standard classical field (SGPE), blue: rSGPE (matches the exact result). 
Yellow: region within the optimal cutoff \cite{Pietraszewicz15}.
}
\label{fig:id-unif}
\end{figure}

\begin{figure}
\begin{center}
\includegraphics[width=\columnwidth]{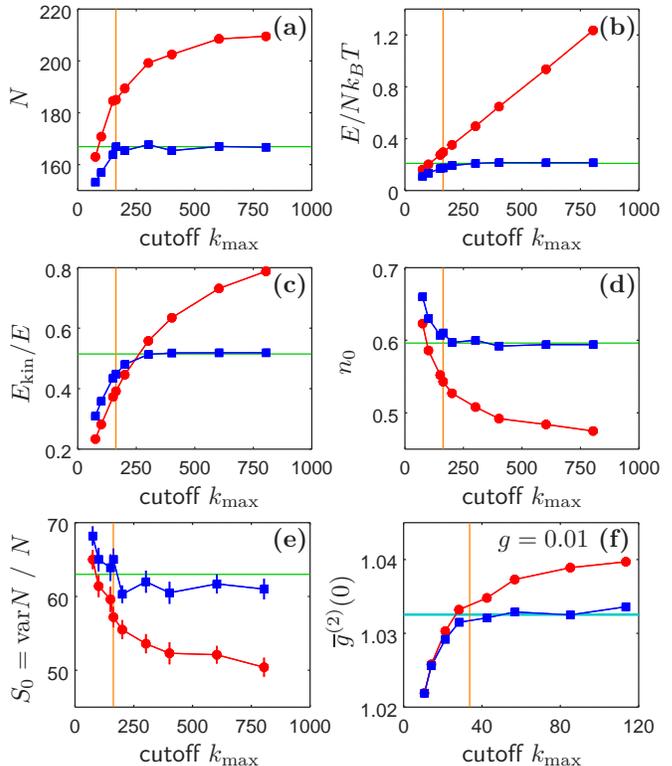}
\end{center}
\vspace*{-0.5cm}
\caption{
Cutoff dependence of observables for 1d gases in a trap. Panels (a-e): ideal, (f): interacting. Grand canonical ensembles, calculated using the standard SGPE (red) and regularized rSGPE \eqn{rSGPE} (blue). Exact solutions in green,
extended Bogoliubov in cyan \cite{Mora03}. The optimum cutoff of $\ve_{\rm cut} = \hbar^2 k_{\rm max}^2/2m = 1.34 k_BT$ \cite{Pietraszewicz15}  is shown in orange. Error bars are smaller than symbols if not visible. 
Details for (a-e) are $\omega=489.9$, $T=10^4$, $\mu=-T/100+\omega/2$, $g=0$, box size $L = 1$. 
Details for (f) are $\omega=1$, $T=428.31$, $\mu=22.41$, $g=0.01$, $L = 113.427$.
Common for all:  $\hbar=m=k_B=1$,  $\gamma=0.1$, $M=k_{\rm max}L/\pi$, $10^4$ trajectories, $M_{\beta}=1$, $\Omega_{\rm cap}=3$.
}
\label{fig:idtrap}
\end{figure}

We perform the next check on a uniform ideal gas. This demonstrates the fundamental difference in how energy and particles are distributed in each method.
Fig.~\ref{fig:id-unif} shows the (kinetic) energy held in each mode, which follows from taking $\hbar\omega=\hbar^2|\bo{k}|^2/2m$ in \eqn{NU=0}.
For the SGPE the equipartition of energy among modes characteristic for the UV catastrophe appears. 
The yellow area shows how choosing a cutoff (in this case from \cite{Pietraszewicz15}) tries to deal with this.
The integrated total energy agrees with the value from full quantum mechanics, but it is distributed in an artificial way.
All these problems are avoided by the regularized theory.

Consider now the trapped 1d gas. We first test the ideal gas, which is a nontrivial system for semiclassical methods.  
Later, we test an interacting quasicondensate.
Fig.~\ref{fig:idtrap} shows the cutoff dependence of the following observables:
Total particle number $N$;
Energy per particle $E/N$;
Fraction of energy that is kinetic $E_{\rm kin}/E$; The
condensate fraction $n_0$ calculated according to the Penrose-Onsager criterion, i.e. the largest eigenvalue of the one-body density matrix $\langle \phi(\bo{x})^*\phi(\bo{x}')\rangle$;
The static structure factor $S_0 = {\rm var} N / N$ describing the density grains in the gas \cite{PDJPYang17}; 
The integrated density correlation $\wb{g}^{(2)}(0)=(L/N^2)\int dx\ \langle|\phi(x)|^4\rangle$.
In the standard classical field treatment 
the UV divergence rears its head:
Many observables do not converge as the lattice is made finer ($k_{\rm max}$ grows), or converge to incorrect values. Optimal cutoffs depend on the observable in question.

\begin{figure}
\begin{center}
\includegraphics[width=\columnwidth]{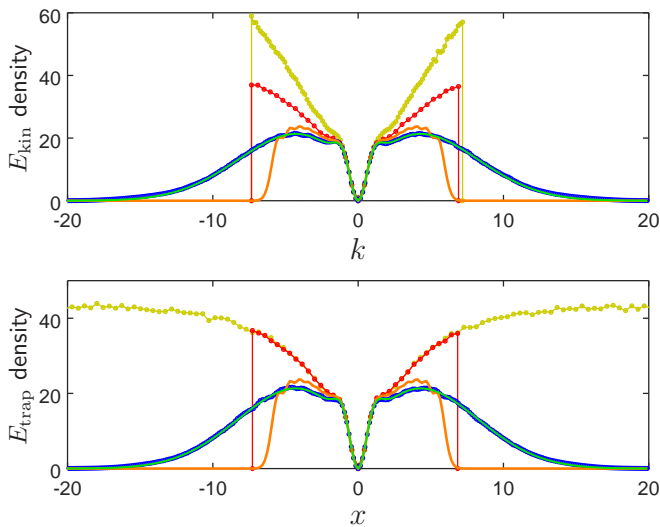}
\end{center}
\vspace*{-0.5cm}
\caption{
Kinetic and trap energy densities 
in the thermal equilibrium state of regularized and standard classical field theory. 
Blue: rSGPE simulation, green: exact quantum mechanical values. 
Warm colours: standard classical fields. 
Orange: using a harmonic oscillator basis with the cutoff $E_c=k_BT$ from \cite{Witkowska09}, optimized for condensate fraction. 
Red: using a plane wave basis with the LDA optimized cutoff  $\ve_{\rm cut}=\hbar^2 k_{\rm max}^2/2m=1.34k_BT$ from \cite{Pietraszewicz15}, and optimal box size $L=2\hbar k_{\rm max}/m\omega$ as per \cite{Bradley05}.
This gives good single particle energies. 
Yellow: in a plane wave basis with $\ve_{\rm cut}=1.34k_BT$ but a wide box $L$, the same as for the rSGPE. 
System parameters in trap units: $\omega=1$, $T=20.106$, $\mu=-T/100+\omega/2$, $L=40.1$ (except red).
Numerical details:  $10^4$ trajectories, $M=256$ points, $t_{\rm max}=200$, $\Delta t=0.0032$, $M_{\beta}=1$, $\Omega_{\rm cap}=4$. 
}
\label{fig:trapideal}
\end{figure}

Fig.~\ref{fig:idtrap} (a-e) concern the ideal gas and display the exact quantum predictions in green, the results of the  standard SGPE in red, 
and of the regularized equation \eqn{rSGPE} in blue.
The improvement from standard SGPE to rSGPE is tremendous. Notably, 
\begin{enumerate}
\item The UV divergence seen in the standard calculation is vanquished completely --- the rSGPE data stabilize to an asymptotic value as cutoff grows. 
\item 
The values they converge to, are in fact the exact quantum ones. 
\item Stabilization occurs at cutoff energy around $4k_BT$. 
\end{enumerate}

The source of the remarkable accuracy of the rSGPE lies in its match to the exact energy densities $E_{\rm kin}(k) = n(k)\hbar^2k^2/2m$ and $E_{\rm trap}(x) = n(x)m\omega^2x^2/2$, as
 shown in Fig.~\ref{fig:trapideal}.
In contrast, the standard c-field calculations all have major flaws despite optimized parameters.  
The trap basis calculation (orange) best matches the local energy density, but overall energy is much underestimated. 
The plane wave calculation (red) matches total energy but overestimates its local density. The yellow data are for a larger box and match neither.

The test results for an interacting thermal trapped quasicondensate in 1d are also satisfying (Fig.~\ref{fig:idtrap}(f)\,).  
The density correlation $\wb{g}^{(2)}(0)$, which is always trivially 2 in the ideal gas, takes on nontrivial values when interaction is present. 
The SGPE suffers from a UV divergence, as expected, 
while the regularized calculation again obtains the correct values. They agree with the extended Bogoliubov result \cite{Mora03}, which is accurate when $g^{(2)}(0)$ is close to one.

\section{A demanding trial: Collective mode frequency in 3d}
\label{JILA}

\subsection{Status to date} 

The 1997 JILA experiment  \cite{Jin97} has long served as a litmus test for the accuracy of dynamical theories as high temperature is approached.
Exciting a thermal cloud makes it oscillate at twice the trap frequency $\nu_r$, whereas the $m=0$ quadrupole collective mode of a pure condensate had frequency $1.85\nu_r$. 
The experiment determined that there is a sudden increase in the condensate oscillation frequency up to $2\nu_r$ around $T\approx 0.7T_c$. 
It is attributed to increasing drag from the thermal cloud
\cite{Morgan03}.
The body of theory surrounding the topic is well described in \cite{Proukakis08}, although most numerical methods could not replicate the full behavior.

One c-field attempt, \cite{Bezett09a}, did not predict a rise in frequency at all, while  \cite{Karpiuk10} saw it only at a temperature that was noticeably too high ($0.8T_c$). 
The only simulation that achieved a rough match to the experiment was made using ZNG theory \cite{Jackson02}. 
The second-order Bogoliubov study of \cite{Morgan03}, though not a simulation itself, was able to predict the condensate frequency well. Unfortunately, the last two approaches are 
less versatile, and unable to model any low lying coherent modes besides the condensate \cite{FINESS-Book-Wright}. 

\subsection{Experimental and numerical procedure}
The experimental runs began with preparation of a ${}^{87}$Rb gas in thermal equilibrium in a 
harmonic trap with frequencies $\nu_x = \nu_y = \nu_r = 129$ Hz and $\nu_z=365$ Hz. 
Gases at various temperatures were prepared ranging from $0.4T_c$ to $1.3T_c$. They can be parametrized by a reduced temperature 
\eq{Ti}{
T' = \frac{T}{T_c^0(N)} = \frac{k_BT\,\zeta(3)^{1/3}}{\hbar\omega_{\rm ho} N^{1/3}},
}
relative to the ideal gas critical temperature  $T_c^0\approx T_c$. Here $N$ is the particle number and $\omega_{\rm ho} =2\pi (\nu_x\nu_y\nu_z)^{1/3}$. 
In order to excite collective motion of the cloud, the trap frequencies were modulated sinusoidally with driving frequency $\nu_d$ for 14ms.
After this, the driving was turned off, and the excited cloud allowed to relax in the trap. 
After a relaxation time $t_r$, the cloud was released, and the far field image, i.e. the density integrated over the z direction, was recorded. In fact, the image largely corresponds to the momentum distribution in the cloud at $t_r$ in the x-y directions.  
The widths of the condensate and thermal parts of the cloud, $w_0(t_r)$ and $w_{\rm th}(t_r)$, were determined from bimodal fits. 
An exponentially damped sinusoidal fit was made to these widths, to extract collective mode frequencies $\nu(T')$ and damping rates $\Gamma(T')$ for both condensate and thermal cloud oscillations.

Our simulations followed the experimental procedure.
We ran the rSGPE \eqn{rSGPE} to stationarity to obtain a thermal ensemble of fields $\phi(\bo{x})$ that matched experimental $T$ and $N$. 
After, these fields were evolved according to the experimental protocol using a Gross-Pitaevskii equation, and frequencies extracted. 
This later part of the evolution had no reservoir coupling to avoid undue external damping of the oscillation generated by the driving.
Details of the simulation and data analysis are given in the supplementary material \cite{supp}, Sec.~\ref{S:3D}.

\begin{figure}
\begin{center}
\includegraphics[width=\columnwidth]{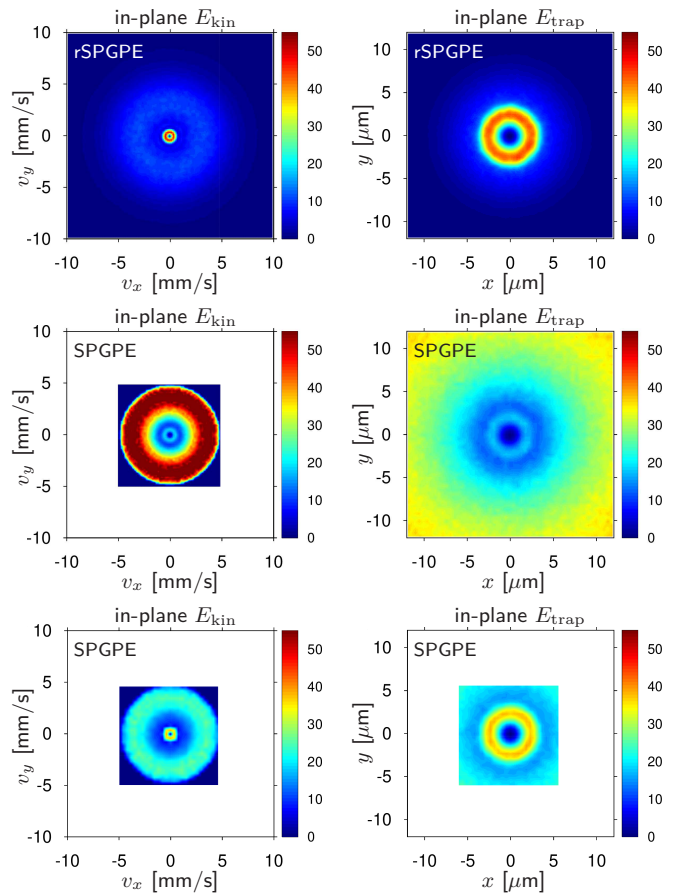}
\end{center}
\vspace*{-0.5cm}
\caption{
Single-particle energy distributions in a trapped 3d gas at thermal equilibrium ($T'=0.428$, $N=6230\pm20$, $T=65$nK) using several methods.  
The left column shows in-plane kinetic energies $n_2(k_x,k_y)\times\hbar^2(k_x^2+k_y^2)/2m$, and the right column in-plane trapping energies $n_2(x,y)\times\pi m\nu_r^2(x^2+y^2)/2$. Both in units of $k_BT/(\mu{\rm m})^2$.
The densities $n_2$ are integrated over the $z$ direction as in experimental images. 
The top row shows rSGPE calculations using a numerical grid as shown ($24.6\times24.6\times8.7\mu$m box, maximum allowed momentum along each axis given by $\hbar^2k_{\rm max}^2/2m=\ve_{\rm cut}=8.16k_BT$). 
The middle row shows SPGPE calculations using the same box, with the optimized energy cutoff \cite{Pietraszewicz15} at $1.91k_BT$. 	
The bottom row uses that same cutoff but the optimized box size from \cite{Bradley05} that matches maximum kinetic and trap energies: $x_j^{\rm max} = \hbar k_{\rm max}/m\omega_j$.	
Chemical potentials $\mu$ were 61.5, 42.6, and 56.4nK, top to bottom.
}
\label{fig:hsp}
\end{figure}

The top row of
Fig.~\ref{fig:hsp} displays the distributions of single-particle energies (kinetic and trap) in the generated thermal state. 
They crisply show two rings: an inner one, due to quantum pressure in the condensate, and the outer one, due to the kinetic energy of particles in the thermal cloud. 

Comparable SPGPE calculations 
using an optimized energy cutoff in k-space at a kinetic energy of $1.91k_BT$ \cite{Pietraszewicz15} are shown in the lower rows. 
The lowest row uses a balanced box size \cite{Bradley05}. 
The cutoff problems here are even stronger than in 1d.
The thermal cloud has an unnatural distorted distribution in both k-space \emph{and} x-space. The trap energy continues to have divergent behavior almost as if there was no cutoff. The inner part of the system is also affected indirectly: While the quantum pressure ring is present, it is weakened and depends strongly on the box size, which affects the condensate fraction.

In contrast, a smooth and complete containment of energy is evident in the regularized ensemble, and there is no dependence on box size. 

\subsection{Collective mode frequencies}
\label{COLL}

\begin{figure}
\begin{center}
\includegraphics[width=\columnwidth]{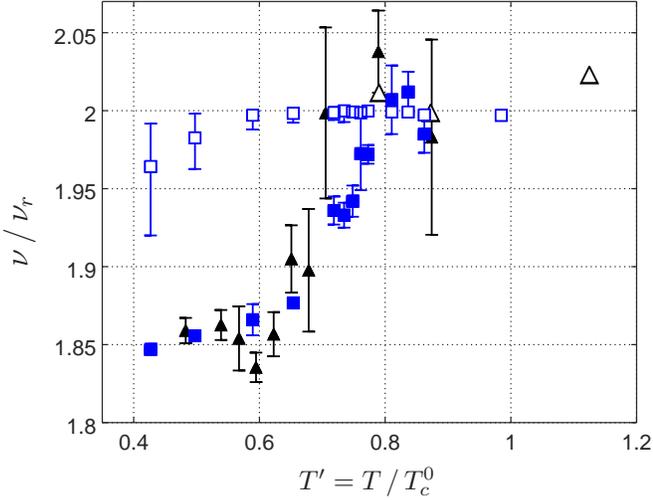}
\end{center}
\vspace*{-0.5cm}
\caption{
Frequencies of the $m=0$ collective mode in the JILA expriment \cite{Jin97}. 
Black triangles: experiment, blue squares: rSGPE simulation. 
Solid symbols: condensate, open symbols: thermal cloud. 
}
\label{fig:freqok}
\end{figure}

\begin{figure}
\begin{center}
\includegraphics[width=\columnwidth]{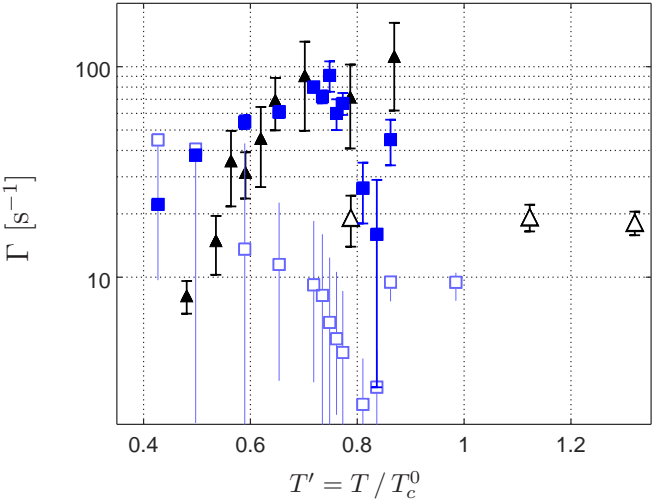}
\end{center}
\vspace*{-0.5cm}
\caption{
Damping rates of the $m=0$ collective mode in the JILA expriment \cite{Jin97}. 
Notation as in Figs.~\ref{fig:freqok} and~\ref{fig:gamok}. 
}
\label{fig:gamok}
\end{figure}

The main numerical results -- the frequencies and damping rates -- are shown in blue in Figs.~\ref{fig:freqok} and \ref{fig:gamok}, respectively. 
Condensate quantities are shown with solid symbols, thermal cloud quantities with open ones. 
The data points are best estimates, while error bars take into account both statistical uncertainty 
and reasonable variation of fitting parameters. 
Details of this 
are given in supplementary material, Sec.~\ref{S:FREQ}  \cite{supp}.

The standout point is that the regularized simulation is finally a classical field treatment that matches the main features seen in the experiment. The frequency changeover for condensate excitations from $1.85\nu_r$ to $2\nu_r$ occurs at exactly the right place, around $0.7T'$. 
Agreement with experimental frequencies is within statistical uncertainty. An exception is the anomalously low experimental data point at $T'=0.6$. 

In the central part of the temperature range  $0.55\lesssim T'\lesssim0.8$ in Fig.~\ref{fig:gamok}, the match of condensate damping is also good. 
The simulation provides new information about the thermal mode frequency and damping for $T'<0.8$, where the experiment had insufficient signal to noise.

\begin{figure}
\begin{center}
\includegraphics[width=0.8\columnwidth]{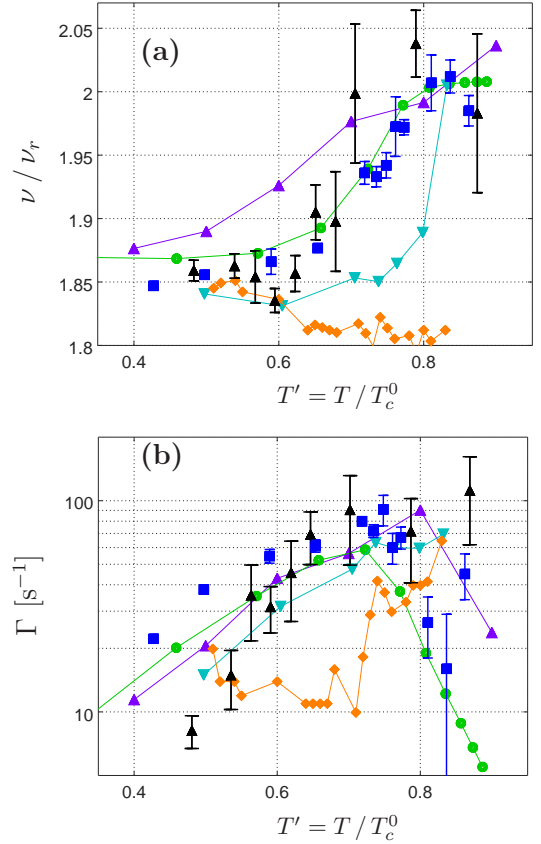}
\end{center}
\vspace*{-0.5cm}
\caption{ 
Comparison of simulations of the $m=0$ collective mode in the condensate. Top: frequency, bottom: damping.  
Violet 
triangles: ZNG simulations \cite{Jackson02}. 
Green circles: 2nd order Bogoliubov \cite{Morgan03}. 
Cyan 
down triangles: classical field with static Hartree-Fock ansatz for the $\mc{I}$ region\cite{Bezett09a}. 
Orange diamonds: classical field with empty $\mc{I}$ region and a somewhat higher cutoff \cite{Karpiuk10}. 
Other notation as in Figs.~\ref{fig:freqok} and~\ref{fig:gamok}. 
Where relevant, the data with the nearest driving frequency to the $\nu_d=2\nu_r$ used for our simulations was chosen. 
}
\label{fig:comp}
\end{figure}

Fig.~\ref{fig:comp} compares condensate results to previous dynamical simulations. 
We see that, not only does the regularized theory give far more accurate values for frequency and damping than standard classical fields \cite{Bezett09a,Karpiuk10}, 
but also improves on the ZNG simulation \cite{Jackson02}. There is also close agreement between the rSGPE and the 2nd-order Bogoliubov \cite{Morgan03}.

However, our calculated condensate dampings deviate in places from the experimental data similarly to ZNG \cite{Jackson02,Straatsma16} and the 2nd-order Bogoliubov. Namely, a slower reduction at low $T'$ occurs, and a rapid drop above $T'\sim0.8$, where the condensate is small. In these regimes the oscillations of $w_0(t)$ were not very sinusoidal. 
At low $T'$ there was a doubled frequency component noted previously \cite{Bezett09a}, while around $T'=0.85$, the collective response was weak. 
As a result, damping values depended quite a lot on details of the fit. 
The discrepancy may be due to differences in fitting details, since the experiment did not give these in full \cite{Jin97}.

\begin{figure}
\begin{center}
\includegraphics[width=\columnwidth]{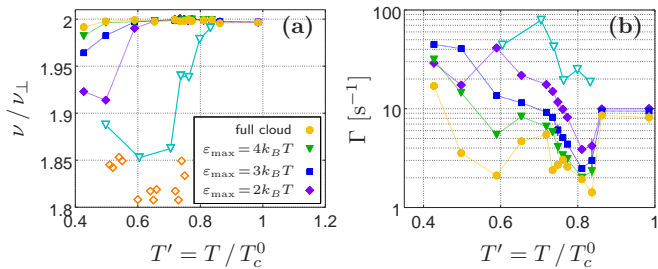}
\end{center}
\vspace*{-0.5cm}
\caption{
Variation of the collective behavior in different parts of the thermal cloud. 
Panel (a) shows the oscillation frequencies, and panel (b) the damping rates. Solid symbols are from the rSGPE. 
The yellow circles come from fits to the entire thermal cloud, while subsequent curves correspond to thermal regions more restricted in momentum to the range $\hbar^2(k_x^2+k_y^2)/2\le\ve_{\rm max}$, as labeled. 
Open symbols come from the PGPE calculations of  \cite{Karpiuk10} (cyan, $E_c\approx2.2k_BT$) and \cite{Bezett09a} (orange, $E_c\approx1k_BT$).  
}
\label{fig:gamthrates}
\end{figure}

Significantly, the regularized theory allows for a quantitative study of the thermal cloud dynamics, which was not possible with earlier theories. 
Fig.~\ref{fig:gamthrates} shows the dependence of the thermal cloud's damping rates and frequencies on the range of momenta included in the analysis. 
The ``full cloud'' data are dominated by the outer tails, whereas the diamond data include only the near tails at kinetic energies below $\ve_{\rm max}=2k_BT$. 
It has been noted \cite{Jackson02} that the thermal gas is not expected to be excited into a true collective mode, but merely a coherent motion of many atoms.
The figure explicitly uncovers this behavior. 
The inner regions damp faster and have a frequency closer to that of the condensate, while the outer ones are almost undamped and continue to oscillate at $2\nu_r$. 
This indicates that the influence of the condensate is responsible for the reduction of the frequency of the thermal cloud below $T'\sim0.5$, and the increase of its damping seen in Figs.~\ref{fig:freqok}-\ref{fig:gamok}.

It also reveals that values found experimentally depended on how much of the thermal cloud emerged above the background noise floor. 
This should be taken into account in future studies of collective modes, and may have played a part in past comparisons.
Accordingly, 
we chose the data points in Figs.~\ref{fig:freqok} and \ref{fig:gamok} from the $\ve_{\rm max}=3k_BT$ curve. They correspond to a noise floor at an occupation of 0.05. 
The uncertainties in Figs.~\ref{fig:freqok}-\ref{fig:gamok}
take into account variation of the $\ve_{\rm max}$ from $2k_BT$ upwards.

Fig.~\ref{fig:gamthrates} additionally shows thermal cloud predictions of the standard classical field from \cite{Karpiuk10,Bezett09a}. 
These past results continue the trend of the rSGPE results as energy $\ve_{\rm max}$ is lowered. 
Notably, they are 
more extreme at $2k_BT$, because they remove 
 all influence of the higher energy tails. 
A more in-depth analysis of the behavior of the collective modes will be reported in a forthcoming paper.

\begin{table*}
\begin{tabular}{|l|c|c|c|c||c|}
\hline
									& ZNG				& \parbox[t]{2.5cm}{PGPE  + Hartree-Fock}		& \parbox[t]{3cm}{classical field / PGPE} 	& \parbox[t]{2.5cm}{SGPE / SPGPE}				& \parbox[t]{2cm}{present work (rSGPE)}
\\
									& \cite{ZNG,Zaremba99,Lee16}
													& \parbox[t]{2.5cm}{\cite{Blakie08,Rooney10,Cockburn11c}} 
																					&\parbox[t]{3cm}{\cite{Kagan97,Goral01,Davis01b,Brewczyk07,Karpiuk10}}
																													& \parbox[t]{3.2cm}{\cite{Stoof99,Duine01,Gardiner03,Proukakis08,Bradley14}} 	&
\\\hline\hline
\parbox[c]{2.5cm}{nonperturbative at low energy}	& \tick			& \tick							& \tick			& \tick			& \tick
\\\hline
many modes in $\mc{C}$						& \cross			& \tick							& \tick			& \tick		 	& \tick
\\\hline
high energy modes							& in $\mc{I}$		& in $\mc{I}$						& \cross			& in reservoir 		& in $\mc{C}$
\\\hline	
dynamics at high energy						& \tick			& \cross							& \cross 			& \cross			& \tick
\\\hline
artificial $\mc{C}$--$\mc{I}$ boundary			& yes \cross		& yes \cross						& yes \cross		& yes \cross		& no \tick
\\\hline
equilibrium temperature						& zero in $\mc{C}$, set in $\mc{I}$ & extracted from $\phi$			& extracted from $\phi$	& set 			& set 
\\\hline
equilibrium ensemble						&  n/a			& microcanonical						& microcanonical		& grand canonical		& grand canonical
\\\hline
cutoff dependence							& no	\tick			& some							& much \cross 		& much \cross		& no \tick
\\\hline
numerical effort							& high 			& medium \tick						& medium \tick 		& medium \tick		& medium \tick
\\\hline
\end{tabular}
\caption{A comparison of the main existing nonperturbative approaches for thermal ultracold gases and the rSGPE method. $\mc{C}$ and $\mc{I}$ refer to the sets of coherent and incoherent modes, respectively.
\label{tab:methods}
}
\end{table*}

Table~\ref{tab:methods} provides a brief comparison to the main existing approaches for simulating thermal Bose gases that are too hot for a Bogoliubov treatment. 

\section{Conclusions}
\label{CONC}

The regularized SGPE \eqn{rSGPE} derived here overcomes the UV catastrophe in the classical wave description of ultracold Bosons,  
and frees it of cutoff parameters. 
A natural and quantum-mechanically correct decay of occupations at high energy is induced. 
This often leaves no arbitrarily chosen parameters, and makes the classical field method quantitative, not merely qualitative, as has often been assumed. 
We have validated the regularized theory for a number of test cases, including the widely known ``hard  problem'' of the $m=0$ collective mode. 
That study let us discover that the properties of thermal clouds observed in experiment have depended on the signal-to-noise ratio.

The rSGPE equation appears to be more versatile than the other methods in Table~\ref{tab:methods}. 
It combines the useful features of both the ZNG and classical field approaches. The dynamics of many highly occupied modes \emph{plus} the thermal cloud can be integrated together. 
The special case of cooling or heating when the system sets its own temperature may be difficult to simulate, because $T$ is set explicitly, just as in the SPGPE.
On the other hand, low temperatures $k_BT\ll\mu$ become accessible.

An important message from the simulations is that there is little sign of error due to a lack of discretization of occupation numbers. 
This has been a major worry for applying c-fields to the almost empty modes above energy $k_BT$. 
A possible explanation is that the influence of these modes on the bulk of the system is primarily through their occupation and fluctuations. Both remain well described, as shown in Sec.~\ref{1MODE}.

Overall, the range of physical phenomena accessible to classical fields widens significantly with the rSGPE. 
In particular, the influence of the thermal modes above $k_BT$ can be studied accurately, and details below the healing length scale become accessible. The latter is crucial for accurate study of superfluid defects.  

An important  practical aspect of the work is the preliminary algorithm described in the Appendix.
This is what allows for a tractable simulation of the otherwise tricky equation \eqn{rSGPE}. 
The final computational effort is comparable to standard SGPE methods, scaling as $M\log M$ in the number of modes, allowing a similar timestep, and not requiring appreciable additional cost for interactions, arbitrary potentials.

Looking widely, regularization of this kind 
is relevant to any models in which the degrees of freedom between the quantum and semiclassical theory differ.  
This includes SGPEs for canonical or other ensembles \cite{Pietraszewicz17b,Rooney12}, and truncated Wigner descriptions of polaritons \cite{Wouters09,Chiocchetta14}, cold atoms \cite{Steel98,Sinatra02,Norrie05,FINESS-Book-Ruostekoski}, or even Yang-Mills theory \cite{Moore97,Tsukiji16}. Truncated Wigner carries the promise of including quantum fluctuations, but requires a more complicated thermal noise term. It may be the next step onward.

\acknowledgments
We are grateful to  many people: Mariusz Gajda, Miros{\l}aw Brewczyk, Emilia Witkowska, 
Matthew Davis, Blair Blakie, Crispin Gardiner, Simon Gardiner, and Nick Proukakis 
for stimulating discussions around this topic.
The work was supported by the National Science Centre (Poland) grant No. 2012/07/E/ST2/01389.

\bibliography{cfields}

\appendix
\renewcommand{\theequation}{A\arabic{equation}}
\setcounter{equation}{0}
\renewcommand{\thesubsection}{A\arabic{subsection}}

\section*{Numerical implementation}
\label{A:ALG}

\subsection{Synopsis}
For a large system, there are two significant issues to face before the equation \eqn{rSGPE} can be integrated efficiently.
\begin{enumerate}
\item The inverse Gibbs factor $\mc{G}=e^{(H-\mu)/T}$ in the decay term is not diagonal in x or k, but a direct matrix representation becomes intractable. 
\item The inverse Gibbs factor $\mc{G}$ becomes very large at high energies, so a straightforward time-stepping algorithm (even a high order one) requires inordinately small timesteps to balance decay with the thermal noise.
\end{enumerate}
Both of the above points turn out to have elegant solutions, but the tricky part is to combine them in an acceptably efficient way. Here this means: keeping the $M\log M$ scaling of numerical effort with the number of modes $M$ that was present in the SGPE. We also stubbornly want to remain in a simple plane-wave basis to preserve generality.  
We find a way to marry these requirements, through the introduction of a setting $\Omega_{\rm cap}$ that controls the amount of computational effort devoted to obtain accurate decay rates. 
Interestingly, and ultimately conveniently, all the above issues occur already in the trapped ideal gas, and no significant computational cost is added by contact interactions.

\subsection{Concept}
\label{A:CONCEPT}

\subsubsection{Trotter decomposition}
One can avoid a matrix implementation of the inverse Gibbs factor $\mc{G}$
by applying a split-step operation in x and k spaces. That is only accurate, however, if the Gibbs factor is close to unity. 
A Trotter decomposition of $\mc{G}$ into $M_{\beta}$ factors of 
\eq{GMB}{
\mc{G}_{M_{\beta}}=e^{(E-\mu)/(M_{\beta}k_BT)},
}
can be used to ensure this. The decomposition leads to $\mc{G}\,\phi = \mc{G}_{M_{\beta}}\cdots\mc{G}_{M_{\beta}}\phi$.

To make the split step, the energy functional $E(\phi)$ can be split into two parts
\eq{Hkx}{
\ve = \frac{\hbar^2|\bo{k}|^2}{2m}, \quad\text{and}\quad H_{\rm x}(\phi) = V(\bo{x}) -\mu + g|\phi(\bo{x})|^2,
}
which are local in k-space or x-space, respectively, such that $E(\phi) = \ve+H_{\rm x}$. Both of these have an associated partial Gibbs factor:
\eq{Gibbsxk}{
\mc{G}_{\rm k}(\bo{k}) = e^{\ve/k_BT}, \qquad\text{and}\qquad \mc{G}_{\rm x}(\bo{x}) = e^{H_{\rm x}/k_BT}.
}
These local factors can then act directly on the field $\phi$ to evaluate $\mc{G}_{\rm x;k}\phi$ with linear cost in $M$. 
A fast Fourier transform $\mc{F}$, which moves between x and k space costs $M\log M$ operations. 
Hence, each Trotter step can be done with cost $2M(1+\log M)$ via
\eq{Trotterstep}{
\mc{G}_{M_{\beta}}\,\phi(\bo{k}) \approx 
\mc{G}_{\rm k}^{1/2M_{\beta}} \mc{F}\left\{ \mc{G}_{\rm x}^{1/M_{\beta}} \mc{F}^{-1}\left[ \mc{G}_{\rm k}^{1/2M_{\beta}}\phi(\bo{k})\right]\right\},
}
where operators act on the right.  
The symmetric form in \eqn{Trotterstep} is more accurate by an extra order of $\delta\beta=1/M_{\beta}k_BT$ than the non-symmetric one. 
Some factors can be amalgamated due to $\mc{G}_{\rm k}^{1/2M_{\beta}}\mc{G}_{\rm k}^{1/2M_{\beta}}=\mc{G}_{\rm k}^{1/M_{\beta}}$. At the end, the total cost to evaluate $\mc{G}\phi$ is proportional to $2M(1+\log M)M_{\beta}$
(instead of $M^2$ for direct matrix multiplication).

\subsubsection{Noise-decay balance}
For a local process, balance with the thermal noise can be obtained by simply solving a linearized equation in $\phi$, $d\phi/dt = -K\phi + A\eta(t)$ with white noise $\eta(t)$. 
Its solution 
\eq{stochsol}{
\phi(t+\Delta t) = e^{-K\Delta t}\phi(t) +\! \sqrt{\frac{\Delta t}{2K}\left(1-e^{-2K\Delta t}\right)}\,A\,\eta(t)
}
gives an accurate time step $\Delta t$ provided the coefficients $K$ and $A$ have not changed much over the time interval $\Delta t$.
See supplementary Sec.~\ref{S:ALG:CHAL1} for details \cite{supp}. 
This procedure does not require the usual  $K\Delta t\lesssim1$ condition that would appear for time-stepping methods based on Taylor expansions such as Euler or even Runge-Kutta methods. Using this trick frees one from the debilitating exponential condition $\Delta t\ll (\hbar/\gamma k_BT)e^{-H/k_BT}$ that would otherwise appear in attempts to integrate \eqn{rSGPE}.

\subsubsection{Diagonal parts and remainder $R$}
If one could separate the evolution \eqn{rSGPE} into parts diagonal in x and k space, 
then each part could be balanced with the noise in the convenient way that was presented in \eqn{stochsol}, and integration of the equation would be relatively straightforward. Unfortunately, $\mc{G}\phi$ is not cleanly separable into two such diagonal pieces because $\ve$ and $H_{\rm x}$ do not commute.
Nevertheless, 
utilization of a split-step method  to some degree is clearly called for. 
We separate out as much local evolution in x and k space as possible, and deal with the leftover differently. 
One can write the decay term as 
\eq{Gsplit}{
-\frac{\gamma k_BT}{\hbar}\ (\mc{G}-1) = \Gamma_{\rm k} + \Gamma_{\rm x} + R,
}
where
\eq{Gammak}{
\Gamma_{\rm k}(\bo{k}) = \frac{\gamma k_BT}{\hbar}\left(e^{\ve/k_BT}-1\right),
}
\eq{Gammax}{
\Gamma_{\rm x}(\bo{x}) =  \frac{\gamma k_BT}{\hbar}\left(e^{H_{\rm x}/k_BT}-1\right),
}
and $R$ is a nonlocal leftover. It is relegated to being added on in the x-space part of the split-step. 
Notably,  \mbox{$R\approx -\gamma\ve H_{\rm x}/\hbar k_BT$} tends to zero as the linearized limit of the SGPE is approached.

\subsubsection{Structure}
The overall framework is a symmetric split-step method that first does $\Delta t/2$ of kinetic-related evolution in k-space, then $\Delta t$ of the x-space evolution, and at the end again $\Delta t/2$ of kinetic-related evolution in k-space.  
To treat both $\Gamma_{\rm x}$ and $\Gamma_{\rm k}$  using \eqn{stochsol}, the thermal noise term is distributed evenly between x-space and k-space  so that it can balance the decay.
The x-space step involves a nonlinear and non-diagonal evolution, so a midpoint iteration is used to stabilize it \cite{Drummond91}.  For the plain GPE the split-step method is symplectic and has been shown to have $\mc{O}(\Delta t)^2$ accuracy after a time $t$ \cite{Javanainen06}. For this, 
one must use the symmetrized form \cite{Feit82}, and the latest available copy of the field as input at all times (as we do below).

We work on a square grid with $M_j$ points in each direction $j=x,y,z$, and periodic boundary conditions in a  box of widths $L_j=M_j\Delta x_j$ (volume $V$, $\Delta v=V/M$ per grid point). This gives plane wave momentum modes in k-space.  The maximum accessible momentum along each axis is $k^{\rm max}_j=\pi/\Delta x_j$, while k-space is accessed via a Fourier transform of the field
\eq{fft}{
\phi(\bo{k}) = \mc{F}\phi(\bo{x})  = \frac{1}{(2\pi)^{d/2}}\int d^d\bo{x}\, e^{-i\bo{k}\cdot\bo{x}}\phi(\bo{x})
}
implemented using a discrete FFT (FFTW) \cite{FFTW05}.

\subsubsection{Capping the remainder}
The actual timestep that must be used, $\Delta t$, is typically constrained by either the need for the coefficients in \eqn{stochsol} to remain constant (they change due to a relatively slow evolution due to the nonlinear interaction term $g|\phi|^2$)  
or the need for $R\Delta t$ to be small. This last condition actually poses the main problem, since in the highest energy regions of phase-space, $\mc{G}\sim\Gamma_{\rm k}\Gamma_{\rm x}$, and $R\propto\mc{G}$ can be extremely large. 

Fortunately, an accurate depiction of $R$ in the very high energy regions is also unimportant. These are the far tails of the density distribution (whether in x or k space), and the occupation here is approximately $1/\mc{G}$. This part of phase space is effectively vacuum and has negligible effect on anything that is going on in the main part of the system. In fact, just the diagonal decay due to $\Gamma_{\rm x}$ or $\Gamma_{\rm k}$ is enough to make occupations negligible in the regions where $R$ is large. 

To obtain a tractable simulation, we introduce a setting $\Omega_{\rm cap}$ that limits the large values of $R$. 
This cannot be done directly on $R$ because it is a huge matrix which can not be dealt with tractably. 
Instead, we flatten the \emph{partial} Gibbs factors 
\eq{tanh}{
\mc{G}'(\mc{K}) = e^{\Omega_{\rm cap}}\ {\rm tanh}\left[\frac{e^{\mc{K}/k_BT}}{e^{\Omega_{\rm cap}}}\right],
}
with energy argument $\mc{K}=\ve$ or $\mc{K}=H_{\rm x}$.
These $\mc{G}'$ are attenuated once  $\mc{K}$ nears $\Omega_{\rm cap}k_BT$. 
They are used to approximate the full $R$ from \eqn{Gsplit} by 
\eq{RR}{
R\to R' = -\frac{\gamma k_BT}{\hbar}\ \mc{R}',
}
where 
\eqa{dRphi}{
\mc{R}'\phi &=& 
 \sqrt{\mc{G}'\left(\ve\right)}\,\mc{G}'\left(H_{\rm x}(\phi)\right)\sqrt{\mc{G}'\left(\ve\right)}\,\phi \nonu\\
&&-\mc{G}'\left(\ve\right)\,\phi -\mc{G}'\left(H_{\rm x}(\phi)\right)\,\phi + \phi,
}
when $M_{\beta}=1$ and all $\mc{G}'$ act on the right. 
More generally,  
\eqa{dRphiM}{
\mc{R}'\phi &=& 
 [\mc{G}'\left(\ve\right)]^{1/2M_{\beta}}\,[\mc{G}'\left(H_{\rm x}(\phi)\right)]^{1/M_{\beta}}\,\cdots \nonu\\
&&\cdots[\mc{G}'\left(\ve\right)]^{1/M_{\beta}}[\mc{G}'\left(H_{\rm x}(\phi)\right)]^{1/M_{\beta}}[\mc{G}'\left(\ve\right)]^
{1/2M_{\beta}}\,\phi \nonu\\
&&\quad-\mc{G}'\left(\ve\right)\,\phi -\mc{G}'\left(H_{\rm x}(\phi)\right)\,\phi + \phi. 
}
The powers are easily evaluated since all $\mc{G}'$ are local in x or k.
This restricts the values of $\mc{R}'$ to about 
$|\mc{R}'|\lesssim e^{2\Omega_{\rm cap}}$.
Then the timestep becomes restricted only by 
\eq{dtrestrictioncap}{
\Delta t \lesssim  \kappa\ \frac{\hbar}{\gamma k_BT}\ e^{-2\Omega_{\rm cap}}
}
rather than $\ll \frac{\hbar}{\gamma k_BT}\ e^{-2{\rm max}[H]}$. In practice, quite high values of the prefactor $\kappa\sim 1-2$ turned out to be sufficient. 
As a result of \eqn{dtrestrictioncap} and \eqn{tanh},
$\Omega_{\rm cap}$ is a control knob that can be used to increase the accuracy of the leftover at the expense of smaller timesteps. 
It basically marks the energy (in $k_BT$ units) at which remaining inaccuracy starts to appear. 

\subsection{Time step algorithm}
\label{A:STEP}

\textbf{\{1\}} The starting field 
is $\phi(\bo{k},t)$. 
Kinetic evolution by $\Delta t/2$ gives
\eq{phi1}{
\phi_1 = e^{-(i\ve/\hbar+\Gamma_{\rm k})\frac{\Delta t}{2}}\phi + \eta^{(k)}_1(\bo{k},t)\sqrt{\frac{\gamma k_BT\Delta t}{\hbar\Gamma_{\rm k}}\left(1-e^{-\Gamma_{\rm k}\Delta t}\right)},
}
with decay rate \eqn{Gammak}. 
The complex noise $\eta^{(k)}_1(\bo{k},t)$ is generated at each full timestep for each $\bo{k}$ using independent Gaussian random variables of variance $V/(4(2\pi)^d\Delta t)$ for the real part and the same variance for the imaginary part.
The field $\phi_1(\bo{k})$ is then transformed to $\phi_1(\bo{x})$ in x-space by the inverse of \eqn{fft}. 

\textbf{\{2\}} The x-space split step is more involved. It is begun by calculating evolution to the midpoint at $t+\Delta t/2$
\eq{phi2}{
\phi_2 = e^{K(\phi_1)\Delta t/2}\phi_1 +  \left(e^{K(\phi_1)\Delta t/2}-1\right)\frac{C(\phi_1)}{K(\phi_1)} +B(\phi_1)\frac{\Delta t}{2}.
}
The form of the first two terms on the RHS comes from the solution of an equation $\frac{d\phi}{dt} = K\phi + C$ with constant coefficients. This is always more accurate than a simple Euler step. It is particularly relevant when ${\rm Re}[K]\Delta t\lesssim-1$, which tends to happen in the high energy part of the system.
The quantities in \eqn{phi2} are 
\eqa{Kx}{
K(\phi) &=& -iH_{\rm x}(\phi)/\hbar -\Gamma_{\rm x}(\phi),\\
C(\phi) &=& R'\phi = -\frac{\gamma k_BT}{\hbar}\,\mc{R}'\phi\\
\label{Bx}
B(\phi) &=& \eta(\bo{x},t)\sqrt{\frac{\gamma k_BT}{2\hbar\Gamma_{\rm x}\Delta t}\left(1-e^{-2\Gamma_{\rm x}\Delta t}\right)}.
}
The x-space complex noise $\eta(\bo{x},t)$ is also generated at each timestep and for each $\bo{x}$ using independent Gaussian random variables. Their variance is $1/(2\Delta v\Delta t)$ for the real part and the same variance for the imaginary parts.
The $\mc{R}'\phi$ is evaluated using \eqn{dRphi} or \eqn{dRphiM}, which involves $3+2M_{\beta}$ Fourier transforms each time.

\textbf{\{3\}} The x-space sub-step is finished by using $\phi_2(\bo{x})$ to evaluate the derivative for the full step. 
\eq{phi3}{
\phi_3 = e^{K(\phi_2)\Delta t}\phi_1 +  \left(e^{K(\phi_2)\Delta t}-1\right)\frac{C(\phi_2)}{K(\phi_2)} +B(\phi_2)\Delta t.
}
Thus \{2\} and \{3\} realize the semi-implicit midpoint algorithm. 

\textbf{\{4\}} Observables that depend on x-space quantities are calculated at this stage using $\phi_3(\bo{x})$. 

\textbf{\{5\}} The final, k-space sub-step by $\Delta t/2$ is begun by Fourier transforming  $\phi_3(\bo{x})$ to $\phi_3(\bo{k})$. 
Then, 
\eqa{phi3i}{
\phi(\bo{k},t+\Delta t) &=& e^{-(i\ve/\hbar+\Gamma_{\rm k})\frac{\Delta t}{2}}\phi_3(\bo{k},t) \\
&&+ \eta^{(k)}_2(\bo{k},t)\sqrt{\frac{\gamma k_BT\Delta t}{\hbar\Gamma_{\rm k}}\left(1-e^{-\Gamma_{\rm k}\Delta t}\right)}\nonu
}
is made using the a freshly generated set of noises $\eta^{(k)}_2(\bo{k},t)$ with the same statistical properties as $\eta^{(k)}_1(\bo{k},t)$. 

\textbf{\{6\}} Finally, observables that depend on k-space quantities are calculated using $\phi(\bo{k},t+\Delta t)$.

\subsection{Accuracy in the tails}
\label{A:ACC}

\begin{figure}
\begin{center}
\includegraphics[width=0.8\columnwidth]{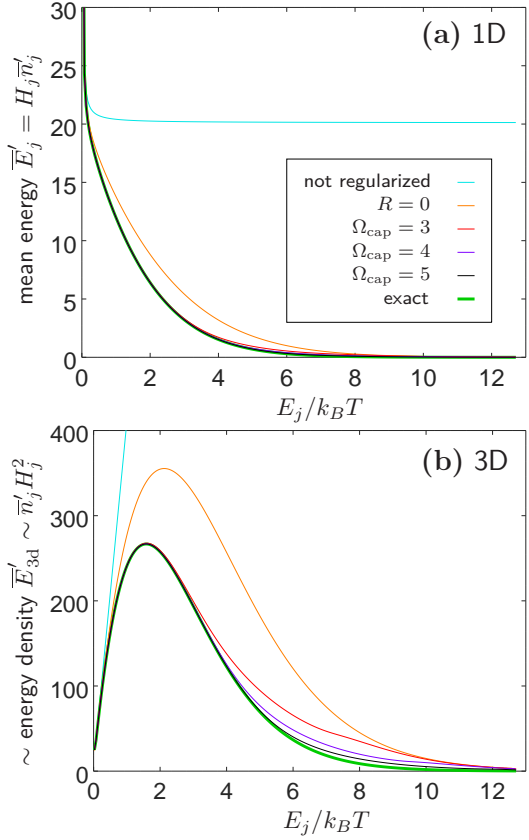}
\end{center}
\vspace*{-0.5cm}
\caption{Energy density in the stationary state of a trapped ideal gas compared between exact and numerical values, depending on the setting $\Omega_{\rm cap}$. 
Panel (a) shows mean energies in eigenstates of $\Gamma$ in a 1d gas with $T=20.1$, whereas panel (b) shows an estimate for the angle-averaged energy density in a 3d gas based on the 1d eigenvalues $\Gamma_j$.  
$E_j$ and $H_j$ are energies of the perfect quantum harmonic oscillator eigenstates, and of the eigenstates calculated on the numerical grid, respectively. They are equal up to $E_j\approx10k_BT$ here. 
}
\label{fig:ecap}
\end{figure}

A good handle on the  accuracy of the simulation is obtained from comparing the actual simulated decay term 
\eq{simGamma}{
\Gamma' = \Gamma_{\rm k} + \Gamma_{\rm x} + R'
}
to the exact one \eqn{Gsplit}. 
Consider the case of zero interactions, which is sufficient for analyzing the tails, which contain the main inaccuracy. 
The eigenvalues of $\Gamma'$ are $\Gamma'_j$, and they can be used to find the stationary values of mode occupations. 
Solution \eqn{stochsol} indicates that the stationary occupations will be $\wb{n}'_j=|\phi|^2\Delta v=\gamma k_BT/\hbar\Gamma'_j$.  
Having these, one can also estimate the mean energies per mode: $\wb{E}'_j=H_j\wb{n}_j$.

The top panel of Fig.~\ref{fig:ecap} shows these $\wb{E}'_j$ for a trapped 1d gas, with a number of $\Omega_{\rm cap}$ settings as they are ramped up. The figure also compares to the exact Bose-Einstein distribution.
For this case, the exact eigenenergies $H_j$ come from a numerical diagonalization of the Hamiltonian.
The bottom panel 
uses the 1d results to estimate the angle-averaged energy density for a 3d gas in a spherical trap, $\wb{E}'_{\rm 3d}$. It assumes that the density of states grows as 
$\sim|\bo{k}|^{d-1}\sim E^{(d-1)/2}$, which is a good estimate far from the ground state. 

Importantly, 
\begin{itemize}
\item[a)] The rudimentary $R=0$ already prevents the UV divergence, though it is not very accurate.
\item[b)] Accuracy improves very rapidly as $\Omega_{\rm cap}$ increases.
\item[c)] For quantitative accuracy, $\Omega_{\rm cap}\ge3$ is necessary (and values of $\Omega_{\rm cap}\approx 4-5$ for accurate energy density in 3d).
\end{itemize}

The simulations in Sec.~\ref{JILA} used $M_{\beta}=1$ and $\Omega_{\rm cap}=3$ unless otherwise stated in Table~\ref{tab:data} or explanatory text. 
For our simulations, no significant difference from $M_{\beta}=1$ to $M_{\beta}=2$ was seen even at the lowest $T'=0.427$ (shown in Table \ref{tab:data}), but it will become important at lower temperatures.

\clearpage

\renewcommand{\thefigure}{S\arabic{figure}}
\setcounter{figure}{0}
\renewcommand{\thetable}{S\arabic{table}}
\setcounter{table}{0}
\renewcommand{\thesection}{S\arabic{section}}
\setcounter{section}{0}
\renewcommand{\thesubsection}{\Alph{subsection}}
\renewcommand{\theequation}{S\arabic{equation}}
\setcounter{equation}{0}
\renewcommand{\thepage}{S\arabic{page}}
\setcounter{page}{1}
\renewcommand{\appendixname}{}
\counterwithout*{equation}{section}	
\setcounter{footnote}{0}

\begin{widetext}
\begin{center}
\textcolor[rgb]{0,0,0}{
{\bf Supplementary material for: \\  
A semiclassical field theory that is freed of the ultraviolet catastrophe
}
\\
\mbox{}\\
J. Pietraszewicz and P. Deuar\\
\textit{Institute of Physics, Polish Academy of Sciences, Al. Lotnik\'ow 32/46, 02-668 Warsaw, Poland}
}\\
\vspace*{2em}
\end{center}

Citation numbers in square brackets refer to the bibliography in the main paper. 
\vspace*{1cm}
\end{widetext}

\section{THE STANDARD METHODS FROM TABLE~\ref{tab:methods}}
\label{S:SPGPE}

The stochastic projected Gross-Pitaevskii equation (SPGPE) \cite{Gardiner03,Blakie08,Bradley14} is:
\eqa{SPGPE}{
i\hbar\frac{d\phi(\bo{x})}{dt} &=& \mc{P}_{\mc{C}}\Bigg\{(1-i\gamma(\bo{x}))\mc{P}_{\mc{C}}\Big[\left(H_{\rm sp} -\mu + g|\phi(\bo{x})|^2\right)\phi(\bo{x}) \Big] \nonu\\
&&\qquad + \sqrt{2\hbar\gamma(\bo{x}) k_BT}\,\eta(\bo{x},t).\Bigg\}
}
The $\gamma(\bo{x})$ is a dimensionless prefactor on the coupling strength between the $\mc{C}$ and $\mc{I}$ modes given by \eqn{Gbar}. 
The $\eta(\bo{x},t)$ are complex white noises.
In practice they are approximated by a pair of real Gaussian random variables of variance $1/(2\Delta t \Delta v)$ in the real and imaginary directions, when time steps are $\Delta t$ and volume elements $\Delta v$. 
The equation \eqn{SPGPE} is ergodic, and relaxes in equilibrium to a grand canonical ensemble \cite{Blakie08}.

The unprojected stochastic Gross-Pitaevskii equation (SGPE) \cite{Stoof99,Duine01,Proukakis08} is:
\eqa{SGPE}{
i\hbar\frac{d\phi(\bo{x})}{dt} &=& (1-i\gamma)\Big[\left[H_{\rm sp} -\mu + g|\phi(\bo{x})|^2\right]\phi(\bo{x}) \Big] \nonu\\
&&\qquad + \sqrt{2\hbar\gamma k_BT}\,\eta(\bo{x},t).
}
which has been obtained by several routes \cite{Stoof99,Duine01,Gardiner03}. 
It comes from making two additional assumptions:

(a) The subspace $\mc{C}$ consists of plane-wave modes below a certain momentum cutoff $k_{\rm max} = \pi/\Delta x$, and one works on a discretized square numerical grid in space so that 
the projection in \eqn{SPGPE} is removed. This assumes that the upper half of the allowed momentum modes do not significantly contribute to the nonlinear evolution so that aliasing of the nonlinearity can be ignored. 

(b) A constant value of $\gamma$ is taken instead of 
$\gamma(\bo{x})$. 

A separate class of c-field methods abstains from including the $\mc{I}$ modes in any form. These are the projected Gross-Pitaevskii equation (PGPE) \cite{Davis01b,Bradley05,Blakie08}
\eq{PGPE}{
i\hbar\frac{d\phi(\bo{x})}{dt} = \mc{P}_{\mc{C}}\Big[\left(H_{\rm sp} + g|\phi(\bo{x})|^2\right)\phi(\bo{x}) \Big],  
}
and the simplest GPE (or ``classical field method'') \cite{Kagan97,Goral01,Berloff02,Brewczyk07,Karpiuk10}
\eq{GPE}{
i\hbar\frac{d\phi(\bo{x})}{dt} =[\left(H_{\rm sp} + g|\phi(\bo{x})|^2\right)\phi(\bo{x})
}
in which plane-wave modes are taken and the projection removed. 
These methods are also ergodic due to the nonlinearity and relax to a canonical ensemble.
A fully static contribution of the $\mc{I}$ modes can still be added \emph{post factum} by hand as an ideal gas or in the Hartree-Fock approximation \cite{Blakie08,Rooney10,Cockburn11c}.

Finally, the Zaremba-Nikuni-Griffin (ZNG) theory \cite{ZNG,Zaremba99,Straatsma16,Lee16} includes only a single condensate mode in $\mc{C}$. However, its dynamics is described in detail by a combination of the PGPE \eqn{PGPE} plus coupling terms to the $\mc{I}$ subspace, which contains all other modes. The latter is described using kinetic theory, which is typically implemented with
test particles. This allows for a fully dynamical evolution of both $\mc{I}$ and its coupling to the single-mode condensate but omits nonlinear dynamics that occurs between low energy modes.

\section{CONVERSION FROM MASTER TO STOCHASTIC EQUATIONS}
\label{S:PP}
\subsection{Generalities}
\label{S:PP:GEN}
The procedure (see e.g. \cite{StochMech,QuantumNoise,Drummond80}) relies on the 
correspondence relations between operators and derivatives acting on the kernel $\op{\Lambda}$ defined in \eqn{Lambda}. Namely:
\eqs{identity}{
\op{a}_j\op{\Lambda} &=& \alpha_j\ \op{\Lambda},\\
\dagop{a}_j\op{\Lambda} &=& \left[\beta_j+\frac{\partial}{\partial\alpha_j}\right]\op{\Lambda},\\
\op{\Lambda}\,\dagop{a}_j &=& \beta_j\ \op{\Lambda},\\
\op{\Lambda}\,\op{a}_j &=& \left[\alpha_j+\frac{\partial}{\partial\beta_j}\right]\op{\Lambda}.
}
Let us define a vector of variables $\vec{v} = [\vec{\alpha}, \vec{\beta}] = [v_1,\dots,v_{n},\dots,v_{2M}]$ as a shorthand. 
The identities \eqn{identity} can be used to equate the master equation \eqn{masterG} with integrals of the form
\eqa{integral}{
\lefteqn{\int d^{4M}\vec{v}\ \frac{\partial P(\vec{v})}{\partial t}\op{\Lambda}(\vec{v}) = 
\int d^{4M}\vec{v}\ P(\vec{v}) \Bigg\{ }&&\\
&&\qquad\qquad C(\vec{v}) + \sum_{v\in\vec{v}} A_v(\vec{v}) \frac{\partial}{\partial v}
+\sum_{v,u\in\vec{v}} \frac{D_{vu}(\vec{v})}{2}\frac{\partial^2}{\partial v\partial u}\Bigg\}\op{\Lambda}(\vec{v}),\nonu
}
where $C$, $A$, and $D$ are complex coefficients. Integration by parts of Eq.(\ref{integral}) gives 
\eqa{parts}{
\lefteqn{\int d^{4M}\vec{v}\ \op{\Lambda}(\vec{v})\frac{\partial P(\vec{v})}{\partial t} = 
\int d^{4M}\vec{v}\ \op{\Lambda}(\vec{v}) \Bigg\{ }&&\\
&&\qquad\qquad
C(\vec{v}) - \sum_{v\in\vec{v}} \frac{\partial}{\partial v}A_v(\vec{v})
+\sum_{v,u\in\vec{v}} \frac{\partial^2}{\partial v\partial u}\frac{D_{vu}(\vec{v})}{2}\Bigg\}P(\vec{v}),\nonu
}
provided boundary terms are zero (as should occur if the distribution is well behaved). 
The coefficients $C$ always end up summing to zero for a master equation.  
Further, one solution of \eqn{parts} is simply that the integrands equal. This gives a Fokker-Planck equation (FPE)
\eq{FPE}{
\frac{\partial P}{\partial t} = \left\{-\sum_{v\in\vec{v}} \frac{\partial}{\partial v}A_v(\vec{v}) +\sum_{v,u\in\vec{v}} \frac{\partial^2}{\partial v\partial u}\frac{D_{vu}(\vec{v})}{2}\right\}P.
}
Using vector notation $\vec{w} = [\vec{v}^{\prime}, \vec{v}^{\prime\prime}]$ for real and imaginary parts 
 of $v_j$, one can rewrite the FPE as:
\eq{FPEr}{
\frac{\partial P}{\partial t} = \left\{-\sum_{w\in\vec{w}} \frac{\partial}{\partial w}\wb{A}_w(\vec{w}) +\sum_{w,z\in\vec{w}} \frac{\partial^2}{\partial w\partial z}\frac{\wb{D}_{wz}(\vec{w})}{2}\right\}P.
}
Samples of the $P$ 
distribution then evolve according to 
\eq{stochr}{
\frac{dw}{dt} = \wb{A}_w(\vec{w}) + \sum_n \wb{B}_{wn}(\vec{w}) \xi_n(t),
}
where $\xi_n$ are independent white real noises  of mean zero and variance $\langle\xi_n(t)\xi_n'(t')\rangle = \delta_{nn'}\delta(t-t')$. 
 The noise indices $n$ are not necessarily variables in $\vec{w}$.
The real noise matrix $\wb{B}$ is a decomposition of $\wb{D}$  that obeys $\wb{D}=\wb{B}\wb{B}^T$.
Conversion to \eqn{stochr} is only possible if the real diffusion matrix $\wb{D}$ has no negative eigenvalues.

The positive-P kernel $\op{\Lambda}$ is set up to be analytic in complex variables $\alpha$ and $\beta$ so that the decomposition is
always made possible. 
An appropriate juggling of the derivatives $\partial\op{\Lambda}/\partial{\alpha_j} = \partial\op{\Lambda}/\partial{\alpha^{\prime}_j} = -i\partial\op{\Lambda}/\partial{\alpha^{\prime\prime}_j}$ and so on, allows one to obtain
a nonnegative diffusion matrix. It has elements
\eqs{Dr}{
\wb{D}_{v^{\prime},u^{\prime}} &=& \sum_n {\rm Re}[B_{vn}]\ {\rm Re}[B_{un}]\\
\wb{D}_{v^{\prime\prime},u^{\prime\prime}} &=& \sum_n {\rm Im}[B_{vn}]\ {\rm Im}[B_{un}]\\
\wb{D}_{v^{\prime},u^{\prime\prime}} &=& \wb{D}_{v^{\prime\prime},u^{\prime}} = \sum_n {\rm Re}[B_{vn}]\ {\rm Im}[B_{un}]
}
where\footnote{Satisfying $D=BB^T$ is not a problem like it was for $\wb{D}=\wb{B}\wb{B}^T$, because $D$ and $B$ are complex.}
 $D=BB^T$, and also 
$ \wb{A}_{v^{\prime}} = {\rm Re}[A_v],
\wb{A}_{v^{\prime\prime}} = {\rm Im}[A_v].$
Then, after collecting together real and imaginary parts,  the resulting stochastic equations for the complex variables are
\eq{stoch}{
\frac{dv}{dt} = A_v(\vec{v}) + \sum_n B_{vn}(\vec{v})\, \xi_n(t).
}

\subsection{Interaction term}
\label{S:PP:INT}
The Hamiltonian interaction terms in the master equation \eqn{masterG} containing $g$ lead to the 
nonzero coefficients:
\eqs{hamtermsa}{
A_{\alpha_j} &= -&\frac{ig}{\hbar}\int d^d\bo{x}\,\psi^*_j(\bo{x})\phi(\bo{x})^2\wt{\phi}^*(\bo{x}),\\
A_{\beta_j} &=& \frac{ig}{\hbar}\int d^d\bo{x}\,\psi^*_j(\bo{x})\wt{\phi}^*(\bo{x})^2\phi(\bo{x}),\\
D_{\alpha_j,\alpha_{j'}} &= -&\frac{ig}{\hbar}\int d^d\bo{x}\,\psi^*_j(\bo{x})\psi^*_{j'}(\bo{x})\phi(\bo{x})^2,\\
D_{\beta_j,\beta_{j'}} &=& \frac{ig}{\hbar}\int d^d\bo{x}\,\psi^*_j(\bo{x})\psi^*_{j'}(\bo{x})\wt{\phi}^*(\bo{x})^2.
}
Discretizing space on a fine numerical grid with $M_x$ sites and with volume $\Delta v$, 
one can choose the decomposition
\eqs{Bxj}{
B_{\alpha_j,n} &=& \sqrt{\frac{-ig\Delta v}{\hbar}}\, \psi_j^*(\bo{x}_n)\phi(\bo{x}_n),\\
B_{\beta_j,n+M_x} &=& \sqrt{\frac{ig\Delta v}{\hbar}}\, \psi_j^*(\bo{x}_n)\wt{\phi}^*(\bo{x}_n),
}
in which $n=1,\dots,M_x$ enumerates the grid positions $\bo{x}_n$.
Equations directly for the fields $\phi(\bo{x})$ are more useful
than those for $\alpha_j$. Using \eqn{phipp}, \eqn{proj}, \eqn{PCdef}, as well as orthogonality and projector properties of the basis, 
one obtains 
\eqs{stochg}{
\frac{d\phi(\bo{x}_n)}{dt} &=& \sum_{m=1}^{M_x} \Delta v P_{\mc{C}}(\bo{x}_n,\bo{x}_m) \Bigg[ \\
&&-\frac{ig}{\hbar}\phi(\bo{x}_m)^2\wt{\phi}^*(\bo{x}_m) + \sqrt{\frac{-ig}{\hbar\Delta v}}\phi(\bo{x}_m)\xi_m(t) \Bigg]\nonu\\
\frac{d\wt{\phi}(\bo{x}_n)}{dt} &=& \sum_{m=1}^{M_x} \Delta v P_{\mc{C}}(\bo{x}_n,\bo{x}_m) \Bigg[ \\
&&\ \frac{ig}{\hbar}\wt{\phi}(\bo{x}_m)^2\phi^*(\bo{x}_m)  + \sqrt{\frac{-ig}{\hbar\Delta v}}\wt{\phi}(\bo{x}_m)\xi_{m+M_x}(t) \Bigg].\nonu
}
The continuous-space form of these terms is \eqn{ppham}.

\subsection{Nonlinear coupling terms to $\mc{I}$}
\label{S:PP:NL}

The terms in question in the master equation \eqn{masterG}  are:
\eq{nlX}{
\frac{k_BT}{\hbar}\int d^d\bo{x}\,\gamma(\bo{x})\left[ \op{\mc{X}}\op{\rho}\dagop{\phi}-\dagop{\phi}\op{\mc{X}}\op{\rho} + \text{h.c.}\right] 
}
with $\op{\mc{X}}(\op{a}_j) = \left[\mc{G}\circ\op{\phi}\right]^{\S}$. 
Since {\S}  indicates only one-particle processes on $\op{\phi}$, the action of the $\mc{G}$ inside $\op{\mc{X}}$ 
becomes modified by $\S$ to a complex-valued form $\mc{G}_c$, where
\eq{G:}{
\op{\mc{X}}(\op{a}_j) = \left[\mc{G}\circ\op{\phi}\right]^{\S} = \mc{G}_c\circ\sum_{j\in\mc{C}}\psi_j(\bo{x})\op{a}_j.
}
This $\mc{G}_c$
must only extract complex-valued weights $W_j$ from the field to the right as per
\eq{W:}{
\mc{G}_c\circ\sum_{j\in\mc{C}} \op{f}_j(\bo{x}) = \sum_{j\in\mc{C}} W_j\, \op{f}_j(\bo{x}).
}
Using the RHS expression in \eqn{G:}, and \eqn{pprep},
the terms in \eqn{nlX} can be written
\eqa{nlX2}{
&&\frac{k_BT}{\hbar}\int d^d\bo{x}\,\gamma\,\int d^{4M}d\vec{v}\, P\sum_{k\in\mc{C}}\Big\{
\psi_k^*\Bigg[\mc{G}_c\circ\sum_{j\in\mc{C}}\psi_j\op{a}_j\Bigg]\op{\Lambda}\dagop{a}_k \nonu\\
&&\qquad+ \psi_k\op{a}_k\op{\Lambda}\Bigg[\mc{G}_c\circ\sum_{j\in\mc{C}}\psi_j\op{a}_j\Bigg]^{\dagger}
- \psi_k^*\dagop{a}_k\Bigg[\mc{G}_c\circ\sum_{j\in\mc{C}}\psi_j\op{a}_j\Bigg]\op{\Lambda}\nonu\\
&&\qquad- \psi_k\op{\Lambda}\Bigg[\mc{G}_c\circ\sum_{j\in\mc{C}}\psi_j\op{a}_j\Bigg]^{\dagger}\op{a}_k\Big\}.
}
The $\gamma$ and all the  $\psi$ are  $\bo{x}$ dependent. 
Applying \eqn{identity}, one obtains only constant and first derivative terms, because 
due to the property \eqn{W:} of $\mc{G}_c$, the only operator combinations are of
the form
$\op{a}\op{\Lambda}\dagop{a}$, $\dagop{a}\op{a}\op{\Lambda}$ and $\op{\Lambda}\dagop{a}\op{a}$. 
Thus, the only nonzero coefficients in the FPE 
will be
\eqa{nl-coef}{
A_{\alpha_j} &=& -\frac{k_BT}{\hbar}\int d^d\bo{x}'\,\gamma(\bo{x}')\,\psi_j^*(\bo{x}')\left[\mc{G}_c\circ\sum_{l\in\mc{C}}\psi_l(\bo{x}')\alpha_l\right],\nonu\\
A_{\beta_j} &=& -\frac{k_BT}{\hbar}\int d^d\bo{x}'\,\gamma(\bo{x}')\,\psi_j(\bo{x}')\left[\mc{G}_c\circ\sum_{l\in\mc{C}}\psi_l(\bo{x}')\beta_l^*\right]^*.\nonu\\&&
}
Notice that now, after conversion of operators $\op{a}$ and $\dagop{a}$ to complex variables $\alpha$ and $\beta$, the form $\mc{G}_c$ is acting only on complex not operator fields. 
Now, the action of $\mc{G}_c$ on a complex field $f$ is unambiguous, so:
\eq{G:c}{
\mc{G}_c\circ f(\bo{x}) = \exp\left[\frac{H_{\rm sp}+g|f(\bo{x})|^2-\mu}{k_BT}\right]f(\bo{x}) = \mc{G}f.
}
This avoidance of the commutation problem is a consequence of the fact that matters of non-commuting operators have been shunted into the distribution $P$.  
The $A$ coefficients can be reincorporated into the evolution equations of the fields $\phi$ and $\wt{\phi}$. The expressions \eqn{ppGa} are obtained using
\eq{stochG}{
\frac{d\phi(\bo{x})}{dt} = \sum_{j\in\mc{C}} \psi_j(\bo{x}) A_{\alpha_j}\quad;\quad
\frac{d\wt{\phi}(\bo{x})}{dt} = \sum_{j\in\mc{C}} \psi_j(\bo{x}) A_{\beta_j}^*.
}

\section{SINGLE-MODE SOLUTIONS}
\label{S:1MODE}

\subsection{Stationary single-mode solution}
\label{S:PP:1MODE}
The Fokker-Planck Equation corresponding to \eqn{motion1} is 
\eq{FPE1}{
\frac{\partial P}{\partial t} = \frac{\partial}{\partial\alpha}\Big[\alpha\big(i\omega+iUn+\gamma T\mc{G}_1(n)\big)\Big]P + {\rm c.c.} + 
\frac{\partial^2}{\partial\alpha^*\partial\alpha}2\,\gamma\,T P
}
with $n=|\alpha|^2$ and the factor $\mc{G}_1(n) = e^{\frac{\omega-\mu+Un}{T}}$.
Postulating the ansatz
$P(n) = e^{-f(n)/T}$
leads to the condition 
\eqa{cond1}{
nf''(n) &=& \mc{G}_1(n)\left[T+nU-nf'(n)\right] \nonu\\
&&+ f'(n)\left[n-1+\frac{nf'(n)}{T}\right]-T.
}
This equation does not contain $f$, only its derivatives $f' = df/dn$ and $f'' = df'/dn$. 
For a non-interacting system, $f_{\rm id}'= T\left(e^{\frac{\omega-\mu}{T}} - 1\right)$ is a solution, independent of $n$. Comparing the $e^{\frac{\omega-\mu}{T}}$ to the full $\mc{G}_1$, suggests a possible general form:
\eq{cond2}{
f'(n) = T(\mc{G}_1(n) - 1) + \delta f'(n).
}
Substituting \eqn{cond2} into \eqn{cond1}, leads to
$\delta f'(n) = 0.$
Integrating, and normalizing to make $P(n=0)=1$ gives then the solution \eqn{Preg} displayed in the main text.

The form of the full solution \eqn{Preg} is not \emph{a priori} obvious. In the 
ideal gas case it reduces to a familiar Gaussian
\eq{PU=0}{
P_{U=0}(\alpha) = {\rm const}\times \exp\left[-\left(e^{\frac{\omega-\mu}{T}}-1\right)n\right].
}
We can compare the general expression to a naive guess based on substituting the interacting energy functional $\omega+Un/2$ for $\omega$ in \eqn{PU=0}. One finds that 
\eq{Preg-est}{
P(\alpha) \propto \exp\left[-\Big(e^{\frac{\omega-\mu+\frac{Un}{2}}{T}}-1\Big)n - \frac{U^2n^3}{24T^2}e^{\frac{\omega-\mu}{T}}
+\dots \right],
}
which shows that the correction term to the naive exponent is proportional to the square of the spacing of the effective ``energy bands'' $U/T$.

\subsection{Solution for equilibration of the tails}
\label{S:ALG:CHAL1}
Consider the equation:
\eq{a29gen}{
\frac{d\phi}{dt} = -(\Gamma+i\omega)\phi + A\eta_1(t),
}
with $\eta_1(t)$ a complex noise of variance $1/dt$ as in Sec.~\ref{1MODE}, 
Suppose also that $\Gamma$, $\omega$ and $A$ can be assumed constant.
The solution after $\Delta t$ of evolution is 
\be\label{esolgen}
\phi(t+\Delta t) = e^{-(i\omega+\Gamma)\Delta t}\phi(t)
 +A\,\mc{I}(\Delta t)
\ee
with the noise integral
\be
\mc{I}(\Delta t) = \int_0^{\Delta t}\!\!ds\, \eta_1(s)\, e^{-(i\omega+\Gamma)(\Delta t-s)}.
\ee
The solution \eqn{esolgen} is not yet directly useful in this form, because an efficient algorithm should not delve below the time scale of $\Delta t$. However, the statistical properties of $\mc{I}$ are easily found. 
Consider $ds\ll\Delta t$ to take on some tiny but nonzero value.
First note that since $\eta_1(s)$ has random phase, then the distribution of $\eta_p(s) = \eta_1(s) \exp[-i\omega(\Delta t-s)]$ is the same as that of $\eta_1(s)$.
Second, the integral $\mc{I}$ is a sum of Gaussian random variables $v(s) = w(s)\,\eta_p(s)$, where $w(s) = ds \exp[-\Gamma(\Delta t-s)]$. The variables $v(s)$ have variance
\be
\langle|v(s)|^2\rangle = ds\,e^{-2\Gamma(\Delta t-s)}.
\ee
A sum of Gaussian variables $v(s)$ is Gaussian distributed. Hence, $\langle\mc{I}\rangle=0$ and 
\be\label{Ivar}
\langle|\mc{I}(\Delta t)|^2\rangle = \int_0^{\Delta t} ds\ e^{-2\Gamma(\Delta t-s)} = \frac{1-e^{-2\Gamma\,\Delta t}}{2\Gamma}.
\ee
Overall, the solution (\ref{esolgen}) can be written as 
\be\label{esolgen-X}
\phi(t+\Delta t) = e^{-(i\omega+\Gamma)\,\Delta t}\phi(t)  + X(\Delta t),
\ee
where the noise $X(\Delta t)$ is complex Gaussian, and has a variance of 
\be\label{Xvar}
\langle|X(\Delta t)|^2\rangle = \frac{A^2\left(1-e^{-2\Gamma\Delta t}\right)}{2\Gamma}.
\ee
If we compare 
(\ref{Xvar}) with the variance what would be generated purely by the noise field $\eta_1$ over a timestep 
$\Delta t$, then we can rewrite (\ref{esolgen}) as
\eq{esolgen-fin}{
\phi(t+\Delta t) = e^{-(i\omega+\Gamma)\,\Delta t}\phi(t) + A\,\sqrt{\frac{\Delta t}{2\Gamma}\left(1-e^{-2\Gamma\Delta t}\right)}\ \eta_1(t).
}
\eqn{esolgen-fin} leads directly to \eqn{stochsol}, \eqn{phi1}, \eqn{Bx} and \eqn{phi3i}.

\section{THE $m=0$ MODE CALCULATIONS}
\label{S:3D}

\subsection{Phase 1: thermal state}
\label{S:PHASE1}

To obtain a thermal ensemble we start with vacuum $\phi(\bo{x})=0$  and evolve the  rSGPE \eqn{rSGPE} 
with a dimensionless coupling $\gamma=0.1$. 
The numerical grid was chosen so that the accessible single particle energies (kinetic $\hbar k_j^2/2m$ and trapping $V(x) = m\pi\nu_j^2 x_j^2$)
were at least $5k_BT$ in each direction $j=x,y,z$ in space\footnote{Except for the two highest temperature cases, which used $3k_BT$ to reduce computational effort.}.
There were up to $M=3\times10^6$ points on the grid.

Though the Hamiltonian part of the evolution is not in principle necessary for convergence to the thermal state, including it significantly speeds up the equilibration. For our parameters, the speed-up was typically by a factor of $\mc{O}(10)$.
With inclusion of the Hamiltonian part, evolution times of $25/(2\pi\nu_r)$ sufficed to obtain the stationary equilibrium ensemble in most cases. 

\begin{figure}
\begin{center}
\includegraphics[width=0.8\columnwidth]{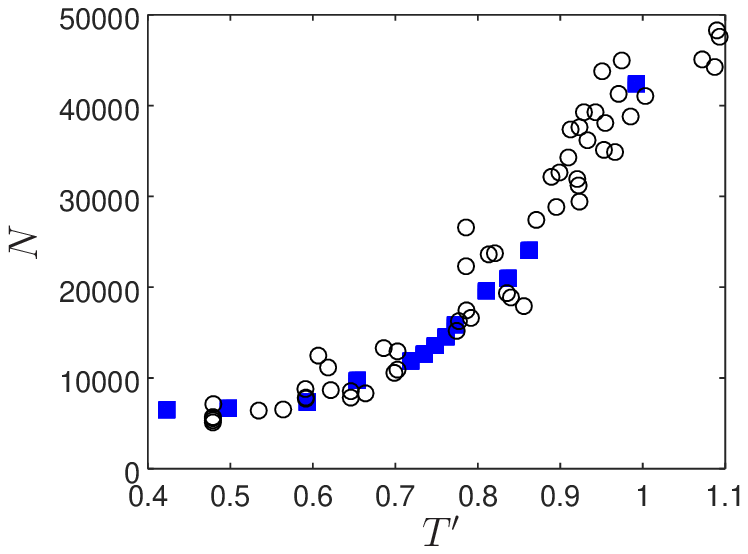}
\end{center}
\vspace*{-0.5cm}
\caption{
Number of particles $N$ used in the simulations of $m=0$ collective oscillations (blue squares) and measured in the experiment \cite{Jin97} (black circles).
}
\label{fig:NTi}
\end{figure}

The atom number $N$ in the experiment was strongly dependent on the temperature. Chemical potentials $\mu(T)$ were chosen 
to match this dependence at each temperature. The resulting $N(T')$ in the experimental and simulated ensembles are shown in  Fig.~\ref{fig:NTi}. 

Most of the data was generated using a Gibbs factor cap setting of $\Omega_{\rm cap}=3$. 
This value gave particle numbers statistically indistinguishable from the exact result for 1d test cases, but
still allowed for reasonable timesteps in 3d -- typically $\Delta t\sim2\mu$s here, so that around $10^4$ steps are needed in each run. 
The value of $\Omega_{\rm cap}=3$ admits minor inaccuracies in the occupations of modes in the far tails above $E_{\rm cap}=\Omega_{\rm cap}k_BT$, as shown in Fig.~\ref{fig:ecap}. We have checked the influence of $\Omega_{\rm cap}$ on our case by
generating
ensembles using the more precise $\Omega_{\rm cap}=4$ or $\Omega_{\rm cap}=5$ for $T=65$nK and $T=135$nK  (see Table~\ref{tab:data}). 
Comparing the occupations and chemical potentials at $T=65$nK, one can see that convergence of accuracy with $\Omega_{\rm cap}$ is rapid because the change from $\Omega_{\rm cap}=4$ to $5$ is much smaller than the one from $\Omega_{\rm cap}=3$ to $4$. 
Using $\Omega_{\rm cap}=3$, a small number of excess particles appear, whose number ranges from 2-3\% of $N$ at low temperatures to 12\% at the highest $T'$. This translates to shifts of -0.005 to -0.03 in $T'$, respectively. 
Such shifts are much smaller than the temperature uncertainty in the experimental data 
(judged to be at the 5-10\% level \cite{Morgan03}). 
A significant dependence on $\Omega_{\rm cap}$ was not seen in frequencies or damping, except for a minor change in $\nu_0$ at $T=135$nK. This is in the steepest part of the curve in Fig.~\ref{fig:freqok}. 
Since $\Omega_{\rm cap}=4$ requires $\sim8$ times smaller timesteps, we remained with the more efficient $\Omega_{\rm cap}=3$ for the majority of temperatures. 
The energy densities in the top row of Fig.~\ref{fig:hsp} were generated using the ultra-precise $\Omega_{\rm cap}=5$, because single-particle energies are the most sensitive quantity to high energy behavior.

\subsection{Phase 2: driving and release}
\label{S:PHASE2}
The system was driven by modulating the trapping potential from $t=0$ until $t_d=14$ms, in order to excite collective motion of the cloud. The modulation was
\eq{omega}{
V(x,t) = \frac{m}{2}\sum_{j=x,y,z} \omega_j^2 x_j^2 \left[ 1+ A_j\cos(2\pi\nu_d t+\theta_j) \right].
}
The $m=0$ quadrupole breathing mode used $A_x=A_y=A$, $A_z=0$, $\theta_j = 0$. As in the experiment and past simulations, amplitudes $A$ were chosen small enough to elicit only a perturbative response in cloud widths. After this driving, free evolution in the unmodulated trap ($A=0$) was continued for another 45ms. 

The data shown in all the figures used a driving frequency of $\nu_d = 2\nu_r$. 
For a few temperatures, we varied this frequency to $1.85\nu_r$ and $1.9\nu_r$ to check for any dependence (Table~\ref{tab:data} lists an example).  We did not find statistically significant differences in response frequencies or damping from the $2\nu_r$ case.

The driving amplitude used for the final analyses varied from $A=0.02$ to $A=0.05$ for the different
values of $T'$. It was chosen to obtain good visibility of the oscillations, while minimizing the excursions of the condensate into larger momenta. 
The resulting response amplitudes were in the range $1\%-8$\% for both
condensate and thermal cloud width. In turn, in the experiment, a 
 nonlinear response for amplitudes was seen 
only above 20\% \cite{Jin97}.

In the free evolution part of the simulation, one wants to study the natural decay rates $\Gamma$. Therefore too-large external damping $\gamma$ is detrimental.  
A test run with the $\gamma=0.1$ used for phase 1 
did show spurious damping of the oscillations in phase 2, compared to $\gamma=0$.
Presumably small enough $\gamma\ll0.1$ would be optimal.
However, we did not want to  introduce additional complexity into an already involved test case.
With a $\gamma=0$ simulation (the plain GPE \eqn{GPE}), we observed that the system was always far from relaxing to the GPE stationary state over the timescales simulated. Moreover, the correct energies and particle numbers are conserved. We therefore expect that the results obtained with the GPE will not veer far from a more careful calculation with nonzero $\gamma$.
This is what was done for the dynamics in phase 2.

\subsection{Extracting frequencies from simulations}
\label{S:FREQ}

\begin{figure}
\begin{center}
\includegraphics[width=\columnwidth]{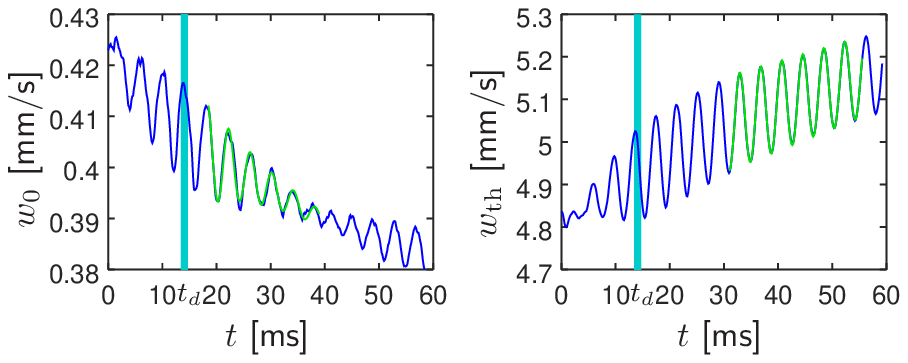}
\end{center}
\vspace*{-0.5cm}
\caption{
Oscillating RMS widths for the condensate (left) and thermal cloud (right), shown in blue. 
Green shows the fit to \eqn{fitfun}.  
The dark cyan line marks $t_d$, the end of driving. 
The system is the $T'=0.748$ case from Table~\ref{tab:data}.
}
\label{fig:beat}
\end{figure}

The data analysis starts from the two-dimensional column density:
\eq{n2}{
n_2(k_x,k_y,t) = \int\!|\phi(\bo{k},t)|^2\, dk_z,
}
averaged over all realizations of the ensemble. This describes the initial velocities in the cloud at release time $t$ and is an estimate for the expanded cloud measured at the detector in the experiment. 
Differences between  $n_2$ and the measured image may arise because of conversion of interaction to kinetic energy during release from the trap \cite{Bezett09a}. However,
this is expected to primarily increase the width of the cloud but not change its oscillation frequency or damping.

The RMS widths of condensate and thermal cloud are obtained from $n_2$, using a procedure inspired by what was done in 1d by Bezett\etal \cite{Bezett09a}. 
We define a momentum magnitude
\eq{r}{
r = \sqrt{k_x^2+k_y^2},
}
and use the disc within $r\le r_0^{\rm max}$ to analyze the condensate. The radius $r_0^{\rm max}$ is chosen to contain the well-defined condensate bulge but not the thermal cloud. 
Its occupation, center-of-mass momentum, and width $w_0(t)$ are  determined by 
\eqa{mk0}{
N_0 &=& \int_{r\le r_0^{\rm min}}\!\!\!\!\!\!\!\!\!dk_xdk_y n_2(k_x,k_y),\nonu\\
\wb{k}_0^{\,x;y}(t) &=& \frac{1}{N_0}\int_{r\le r_0^{\rm min}}\!\!\!\!\!\!\!\!\!dk_xdk_y\ [k_x;k_y]\,n_2(k_x,k_y),\hspace*{-2cm}\\
w_0 = \frac{1}{N_0}&&\hspace*{-0.5cm}\int_{r\le r_0^{\rm min}}\!\!\!\!\!\!\!\!\!\!\!\! dk_xdk_yn_2(k_x,k_y)\!\left[(k_x-\wb{k}_0^{\,x})^2+(k_y-\wb{k}_0^{\,y})^2\right].\nonu
}
For the thermal cloud, the same procedure is followed except that densities in an annulus defined by 
\eq{rth}{
r_{\rm th}^{\rm min}\le r\le r_{\rm th}^{\rm max}
}
are used. This gives the RMS width $w_{\rm th}(t)$.

The inner radii $r_0^{\rm max}$ and $r_{\rm th}^{\rm min}$ were chosen 
to exclude the messy transition region between condensate and thermal cloud.
This region, if present, contains  alternately the tails of the condensate and the inner part of the thermal cloud as they oscillate, and the frequency-doubled oscillations noted by \cite{Bezett09a}. The particular values of $r_0^{\rm max}$ and $r_{\rm th}^{\rm min}$ used for the analysis
vary between ensembles. $\hbar r_0^{\rm max}/m$ was in the range 0.7-0.8mm/s and $\hbar r_{\rm th}^{\rm min}/m$ in the range 0.75-1.5mm/s. 

The outer thermal cloud radius $r_{\rm th}^{\rm max}$ was used to study the energy dependence of the thermal cloud oscillations, as shown in Fig.~\ref{fig:gamthrates}, in which
\eq{rthmax}{
\ve_{\rm max} = \frac{\left(\hbar\,r_{\rm th}^{\max}\right)^2}{2m}. 
}

The widths $w_0(t)$ ad $w_{\rm th}(t)$ are fitted to the function
\eq{fitfun}{
w_{\rm fit}(t) = \wb{w} (1 + \delta w) e^{-\Gamma t} \sin(2\pi\nu t+\theta) + C\,t
}
over a time interval $t_{\rm start} \le t \le t_{\rm end}$. We do this by minimizing the rms deviation between $w_{\rm fit}$ and the actual $w$, while allowing the parameters $\wb{w}$, $\delta w$, $\nu$, $\Gamma$, $\theta$, and $C$ to vary. 
The main results reported in Sec.~\ref{JILA} are $\nu$, $\Gamma$ and their uncertainties. 
Fig.~\ref{fig:beat} shows a typical case out of those for which a clean fit is obtained. 

A complication occurs concerning comparison with the experiment: The dynamics of the widths is considerably less clean than the simple decaying sinusoidal ansatz \eqn{fitfun} used both here and in the experiment (the latter used $C=0$). In particular, beating similar to that reported in \cite{Jackson02} and \cite{Straatsma16} is often seen for $w_0$ (see e.g. Fig.~\ref{fig:beat}). Also, the decay of $w_{\rm th}$ often begins only some time after the cessation of driving, at a time $t_{\rm peak}>t_d$. Moreover, the parameter values and fitting times used in the experiment have apparently been lost \cite{Jackson02}.

We fit the condensate width starting from a time $t_d+2$ms unless the maximum amplitude clearly occurs later. Then,
we start the fit from $t_{\rm peak}$. 
The fitting time extends to the end of the initial decay of amplitude oscillations (due to beating or otherwise). Both timescales match the information provided in the experimental paper. 

For the thermal cloud, it seems reasonable that it
was visible out to energies of $\sim 3k_BT$, so we take the fits with $\ve_{\rm max}=3k_BT$ as the best estimate. 
This is rather uncertain, though, so we also made fits with $\ve_{\rm max}/k_BT = 2, 4, 6, \infty$ and used their spread as a measure of systematic (fitting) uncertainty.
The thermal amplitude peak usually occurs around $t_{\rm peak}\approx30$ms, but is weak so it was not clear if the experiment began fitting at $t_{\rm peak}$ or around $t_d$. Therefore, 
we include fits to both time ranges (from $t_{\rm peak}$ and from $t_{\rm start}=t_d+2$ms) into the range of fitting uncertainties. 

Table~\ref{tab:data} lists the main data obtained from the rSGPE simulations, as well as the primary numerical and fitting settings. 
Statistical uncertainty is estimated by using several ($\mc{S}=6-12$) subensembles of trajectories. Fits are made to each of them. Applying the central limit theorem,
the standard deviation of the subensemble results, divided by $\sqrt{S}$, is the final uncertainty of the mean.
The final uncertainties that are quoted in Table~\ref{tab:data} and shown in the plots combine statistical and systematic uncertainty. The statistical error in the mean is added to the systematic fitting error at both ends of its range.

\begin{turnpage}\begin{table*}
\begin{tabular}{|ll|r@{\ }l|l|l|l|c|cccc|c|c|cc|r|l|}
\hline
\multicolumn{4}{|c|}{Physical parameters}	&\multicolumn{4}{c|}{Collective modes}	&\multicolumn{6}{c|}{Settings}	&\multicolumn{2}{c|}{Fitting times}	& ense-	&	\\
\cline{1-16}
$T$ [nK]\!\!\!\!& $\quad T'$	& $N\qquad$		& $\!\!\!\!\!\mu$ [nK]& $\qquad\nu_0\,/\,\nu_r$	& $\qquad\nu_{\rm th}\,/\,\nu_r$	& $\Gamma_0$ [/s]	& $\Gamma_{\rm th}$ [/s]&  
															$\Omega_{x,y}^{(\rm x)}$ & $\Omega_z^{(\rm x)}$ &$\Omega_{x,y}^{(\rm k)}$& $\Omega_z^{(\rm k)}$ & $\Omega_{\rm cap}$ 
																										& A & $t^0_{\rm fit}$ [ms] & $t^{\rm th}_{\rm fit}$ [ms]	& mble	& remarks\\
\hline\hline
65	& 0.427	& 6318(16)	& 61.5& 1.847(1)		& 1.964(37--66)	& 22.2(7)	& 45(10--46)&  	8	& 8	& 8.16&7.26	&4&	0.03 	& 16--44	& 28--59	& 100	& $r_{\rm th}^{\rm min}=1.5$mm/s	\\	
77	& 0.498	& 6639(10)	& 59	& 1.8557(13)	& 1.983(67--02)	& 38.0(16)	& 41(2--42)	&  	8	& 8	& 8.17&9.20	&3&	0.02	& 16--44 	& 31--59	& 180	& $r_{\rm th}^{\rm min}=1.5$mm/s	\\	
95	& 0.589	& 7497(12)	& 58	& 1.866(10)		& 1.997(88--99)	& 55(4)	& 14(1--43)	&  	8	& 8	& 8.38&7.64	&3&	0.02	& 16--44	& 31--59	& 120	& $r_{\rm th}^{\rm min}=1.5$mm/s    \\	
115	& 0.654	& 9747(15)	& 63	& 1.877(3)		& 1.9983(55--86)	& 61(4)	& 12(3--23)	&  	5	& 5	& 5.15&5.05	&3&	0.02	& 16--44	& 31--55	& 159	&	\\	
135	& 0.719	& 11876(15)	& 62	& 1.936(9)		& 1.9988(41--02)	& 80(4)	& 9(3--19)&  	5	& 5	& 5.38&4.78	&3&	0.02	& 19--44	& 34--55	& 99	&	\\	
141	& 0.735	& 12650(20)	& 61.5 & 1.933(8)		& 1.9999(55--03)	& 72(5)	& 8(2--16)&  	5	& 5	& 4.93&5.55	&3&	0.02	& 19--39	& 31--55	& 66	&	\\	
147	& 0.748	& 13575(20)	& 61	& 1.942(10)		& 1.9992(76--97)	& 91(15)	& 6(2--12)& 	5 	& 5	& 4.92&5.11	&3&	0.02	& 19--39	& 28--59	& 72	& 	\\	
153	& 0.761	& 14537(15)	& 60.5 & 1.973(24)	& 1.9987(63--94)	& 60(10)	& 5(2--11)& 	5	& 5	& 5.17&4.72	&3&	0.02	& 19--42	& 31--55	& 108	&	\\	
160	& 0.773	& 15882(16)	& 60	& 1.972(6)		& 1.9998(75--01)	& 67(8)	& 4(2--8)& 		5	& 5	& 4.85&5.33	&3&	0.02	& 19--39	& 31--55	& 120	&	\\	
180	& 0.811	& 19611(17)	& 53	& 2.007(22)		& 1.9992(8)		& 27(9)	& 2.5(1.0--4.1)&  5	& 5	& 5.38&4.79	&3&	0.05	& 31--52	& 31--55	& 72	& weak response; no clear decay \\	
190	& 0.837	& 20974(15)	& 49	& 2.012(13)		& 1.9992(6)		& 16(13)	& 3.0(1.3--4.5)&	5	& 5	& 5.24&5.44	&3&	0.05	& 49--68	& 16--43	& 72	& weak response; no clear decay \\	
205	& 0.862	& 24060(20)	& 66	& 1.985(12)		& 1.997(2)		& 45(11)	& 9.5(7.7--10.1)&  3	& 3	& 3.07&3.46	&3&	0.05	& 19--47	& 28--59	& 66	&	\\	
285	& 0.985	& 42420(20) & 10	& 			& 1.9970(9)		& 		& 9.5(7.8--10.5)&	3	& 3	& 3.14&3.14 &3&	0.05	& 		& 16--39	& 24	& no condensate \\	
\hline
65 	& 0.429	& 6232(17)	& 61.5& 1.849(23)		& 1.968(67--89)	& 23.2(10)	& 45(12--46)&	8	& 8	& 8.16&7.26	&5&	0.03	& 16--44	& 28--59	& 70	& shown in Fig.~\ref{fig:hsp}	\\ 
65 	& 0.427	& 6301(12)	& 60.2& 1.8509(8)		& 1.960(57--97)	& 26.2(8)	& 54(5--56)	&	8	& 8	& 8.16&7.26	&3&	0.02	& 16--44	& 31--59	& 102	& for comparison of $\Omega_{\rm cap}$	\\	
65 	& 0.429	& 6214(15)	& 60.2& 1.854(3)		& 1.95(1--9)	& 25.8(0)	& 63(5--66) &	8	& 8   & 8.16&7.26	&3&	0.02	& 16--44	& 37--59	& 100	& $M_{\beta}=2$\\	
95	& 0.589	& 7497(12)	& 58	& 1.868(9)		& 1.996(96--01)	& 49(4)	& 16(2--16) &	8	& 8	& 8.38&7.64	&3&	0.02	& 16--44	& 31--59	& 120	& $\nu_d=1.9\nu_r$	\\	
95 	& 0.599	& 7348(15)	& 60	& 1.8600(25)	& 1.998(6--9)	& 46.9(22)	& 15(6--32)	&	5	& 5	& 5.24&5.43	&3&	0.02	& 16--44	& 31--59	& 96	& smaller k-space for comparison\\	
135 	& 0.719	& 11848(17)	& 67	& 1.891(8)		& 1.9995(65--00)	& 78(6)	& 11(4--20) &	5	& 5   & 5.39&4.78	&4&	0.02	& 19--39	& 31--55	& 96	& for comparison of $\Omega_{\rm cap}$\\	\hline
\end{tabular}
\caption{
Details of the collective oscillation data and ensembles analyzed in Sec.~\ref{JILA}. 
The numerical grid is described in terms of the maximum kinetic and trap energies accessible along the axes $j$, in units of $k_BT$, i.e. $\Omega_j^{(\rm k)}=\hbar^2{\rm max}[k_j^2]/(2mk_BT)$
and $\Omega_j^{(\rm x)}=m\omega_j^2{\rm max}[x_j^2]/(2k_BT)$. 
They enter directly into the Bose-Einstein occupations of modes, $N_{BE}\sim1/(e^{\Omega}-1)$, so that a choice of $\Omega=3$ includes practically all density, $\Omega=5$ almost all single-particle energy, and $\Omega=8-10$ any minor remainders. Typical optimum cutoffs for standard classical fields are $\Omega\sim1-2$ \cite{Witkowska09,Pietraszewicz15,Pietraszewicz18a,Pietraszewicz18b}. 
The driving amplitude is $A$, while $t^{0,\rm th}_{\rm fit}$ give the time ranges used for fitting of the best estimate values. 
Time range $t^0_{\rm fit}$ refers to the best condensate estimate, and the range $t_{\rm fit}^{\rm th}$ to the best estimate, $\ve_{\rm max}=3k_BT$, fit of the thermal cloud.
Top block: data shown in Figs.~\ref{fig:freqok}--
\ref{fig:gamthrates}; bottom block: additional cases. 
\label{tab:data}
}
\end{table*}
\end{turnpage}

\end{document}